\documentclass[]{ceurart}

\usepackage{listings}
\usepackage{subcaption}
\usepackage{todonotes}
\usepackage{tikz}
\usepackage{comment}

\begin{document}

\copyrightyear{2024}
\copyrightclause{Copyright for this paper by its authors. Use permitted under Creative Commons License Attribution 4.0 International (CC BY 4.0).}

\conference{EWAF'24: European Workshop on Algorithmic Fairness, July 01--03, 2024, Mainz, Germany}

\title{Exploring Fusion Techniques in Multimodal AI-Based Recruitment: Insights from FairCVdb}

\author[1]{Swati Swati}[%
orcid=0000-0002-7637-6640,
email=swati.swati@unibw.de,
]
\cormark[1]
\address[1]{Research Institute CODE, University of the Bundeswehr Munich, Germany}

\author[1,2]{Arjun Roy}[%
orcid=0000-0002-4279-9442,
email=arjun.roy@unibw.de,
]
\address[2]{Institute of Computer Science, Free University Berlin, Germany}

\author[1]{Eirini Ntoutsi}[%
orcid=0000-0001-5729-1003,
email=eirini.ntoutsi@unibw.de,
]

\cortext[1]{Corresponding author.}

\begin{abstract}
   Despite the large body of work on fairness-aware learning for individual modalities like tabular data, images, and text, less work has been done on multimodal data, which fuses various modalities for a comprehensive analysis. In this work, we investigate the fairness and bias implications of multimodal fusion techniques in the context of multimodal AI-based recruitment systems using the FairCVdb dataset. Our results show that \emph{early-fusion} closely matches the ground truth for both demographics, achieving the lowest MAEs by integrating each modality's unique characteristics. In contrast, \emph{late-fusion} leads to highly generalized mean scores and higher MAEs. Our findings emphasise the significant potential of \emph{early-fusion} for accurate and fair applications, even in the presence of demographic biases, compared to \emph{late-fusion}. Future research could explore alternative fusion strategies and incorporate modality-related fairness constraints to improve fairness. For code and additional insights, visit: \url{https://github.com/Swati17293/Multimodal-AI-Based-Recruitment-FairCVdb}
\end{abstract}

\maketitle

\section{Introduction} 
    The increasing popularity of decision-making algorithms has raised concerns about bias in decision-making, especially towards specific social groups defined by protected attributes such as gender and ethnicity~\cite{raghavan2020mitigating}. Research on fairness-aware learning primarily focuses on individual modalities, such as tabular data~\cite{le2022survey}, text~\cite{cai2022power}, images~\cite{fabbrizzi2022survey}, and graphs~\cite{ghodsi2024towards}. However, there has been less focus on bias in multimodal systems~\cite{booth2021bias}, which can result from integration complexity, unbalanced representation, alignment, and the compounding effect of biases present in each modality. To this end, in this work, we investigate the bias and fairness implications of multimodal AI in automated recruitment systems using the FairCVdb~\cite{pena2020bias} dataset. We use it as a testbed, as it offers diverse data, including images, text, and structured data with intentionally designed gender and ethnicity biases. We focus on fusion techniques for integrating information from different modalities \cite{xue2023dynamic}, specifically analysing \emph{early-} and \emph{late-} fusion techniques known for their straightforward interpretability and widespread usage in multimodal AI systems~\cite{gadzicki2020early,pereira2023comparative}.

\emph{Early-fusion} typically concatenates features from different modalities early on, creating a unified representation of the data \cite{pereira2023comparative}, which simplifies training and effectively captures interactions between modalities \cite{barnum2020benefits}. \emph{Late-fusion,} on the other hand, processes each modality individually before combining their outputs at a later stage, offering flexibility by allowing different processing pathways for individual modalities~\cite{pereira2023comparing}. While \emph{late-fusion} captures modality-specific patterns more accurately, it may overlook lower-level interactions between modalities~\cite{bayoudh2022survey}. By investigating these two fusion strategies, we aim to gain insight into how they impact bias and fairness in automated recruitment processes.

\section{Experimental Setup}
    \noindent \textbf{Dataset:}
The FairCVdb dataset \cite{pena2020bias} comprises of $24,000$ synthetic resume profiles, each featuring demographic characteristics (gender and ethnicity), \emph{textual} data (a short biography), \emph{visual} data (a facial image), and \emph{tabular} data (seven common resume attributes). The resume attributes include occupation, suitability, education, previous experience, recommendation, availability, and language proficiency. Each profile has been generated based on two gender categories and three ethnic categories. The profiles in the dataset are scored based on the likelihood of a candidate being invited to an interview, yielding a numerical score. These scores are assigned either blindly (i.e., without any bias), leading to bias-neutral scores, or with a penalty factor applied to specific individuals within a demographic group, resulting in biased scores. See \cite{pena2023human} for more details. This setup simulates scenarios where cognitive biases, introduced by humans, protocols, or automated systems, influence the decision-making process. 

\noindent \textbf{Evaluation Metrics:} Following \cite{pena2023human}, we use Mean Absolute Error (MAE) to measure prediction error and Kullback-Leibler divergence (KL) to assess demographic bias. For gender, we compare score distributions for males and females; for ethnicity, we perform pairwise comparisons and report the average divergence.

\noindent \textbf{Models:} 
We extend the testbed~\cite{pena2023human} to facilitate multimodal recruitment learning by including  \emph{early-fusion} and \emph{late-fusion} techniques for all three modalities (\emph{textual}, \emph{visual}, \emph{tabular}).

\noindent \textbf{Simulated setups:} We investigated both i) unbiased ideal world setup and ii) real-world setups gender- and ethnicity- biased).

\section{Evaluation Results}
    Figure~\ref{fig:KL-divergence} depicts the KL-divergence (KL), Mean Absolute Error (MAE), and score distributions for gender and ethnicity across different modalities and bias setups. A smaller KL-divergence indicates better alignment between distributions, implying less bias, while a lower MAE indicates a smaller margin of error. Analysis of the score distributions offers additional insights into the models' predictive performance.

    \begin{center}
        
        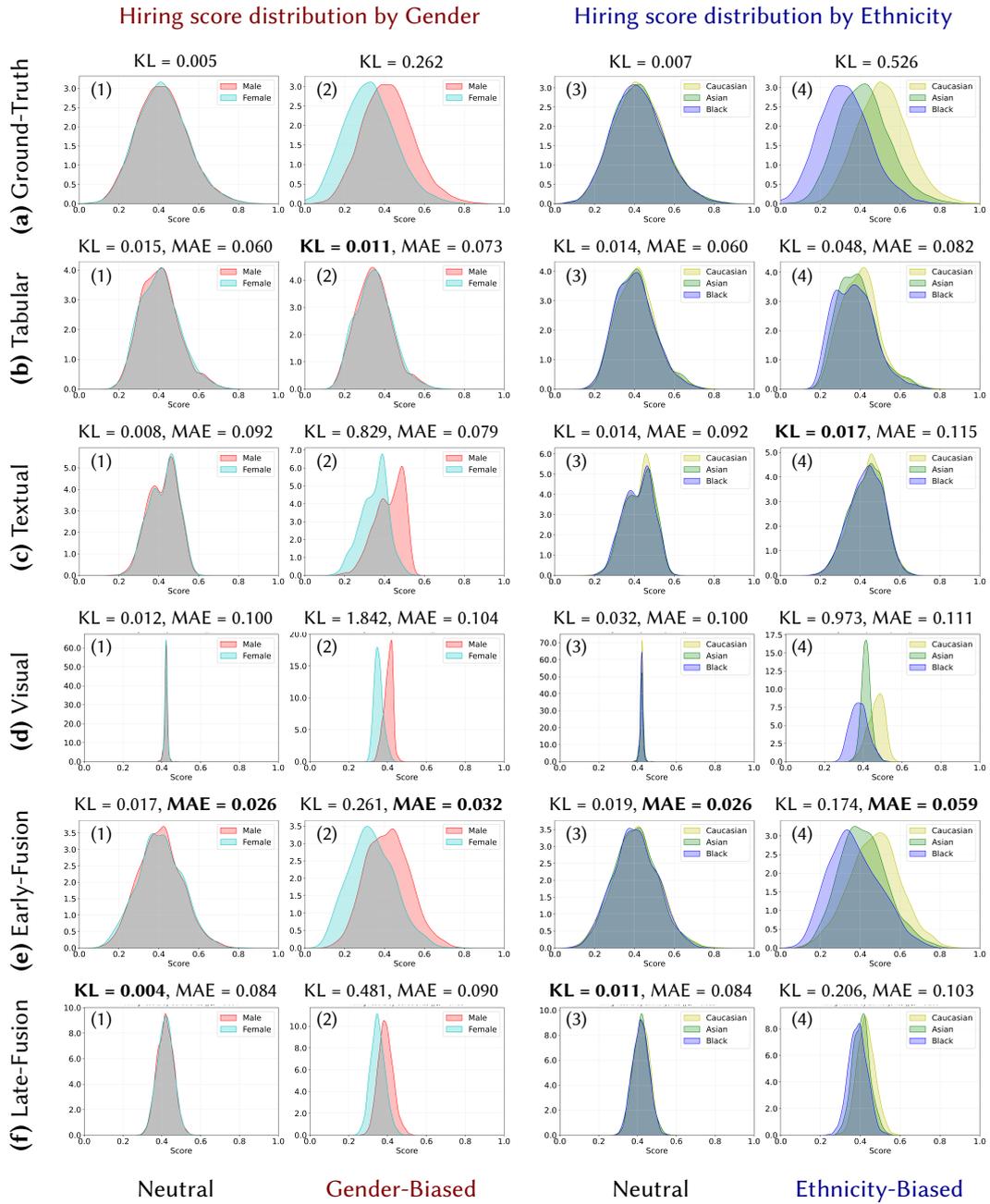
\begin{figure}[!bthp]

            \begin{tabular}{@{}>{\raggedright}p{2cm}@{\hspace{-6\tabcolsep}}c@{\hspace{-0.1\tabcolsep}}c@{\hspace{1.5\tabcolsep}}c@{\hspace{-0.1\tabcolsep}}c@{}}
        
                {} & \multicolumn{2}{c}{\textcolor{Maroon}{Hiring score distribution by Gender}} & \multicolumn{2}{c}{\textcolor{DarkBlue}{Hiring score distribution by Ethnicity}} \vspace{0.2cm} \\


                 \rotatebox{90}{\hspace{-0.1cm}\textbf{(a)} Ground-Truth}
                 
                 & 
                 
                 \begin{tikzpicture}
                    \node[inner sep=0] (image) at (0,0) {\includegraphics[width=0.22\linewidth, trim=5 0 0 24, clip]{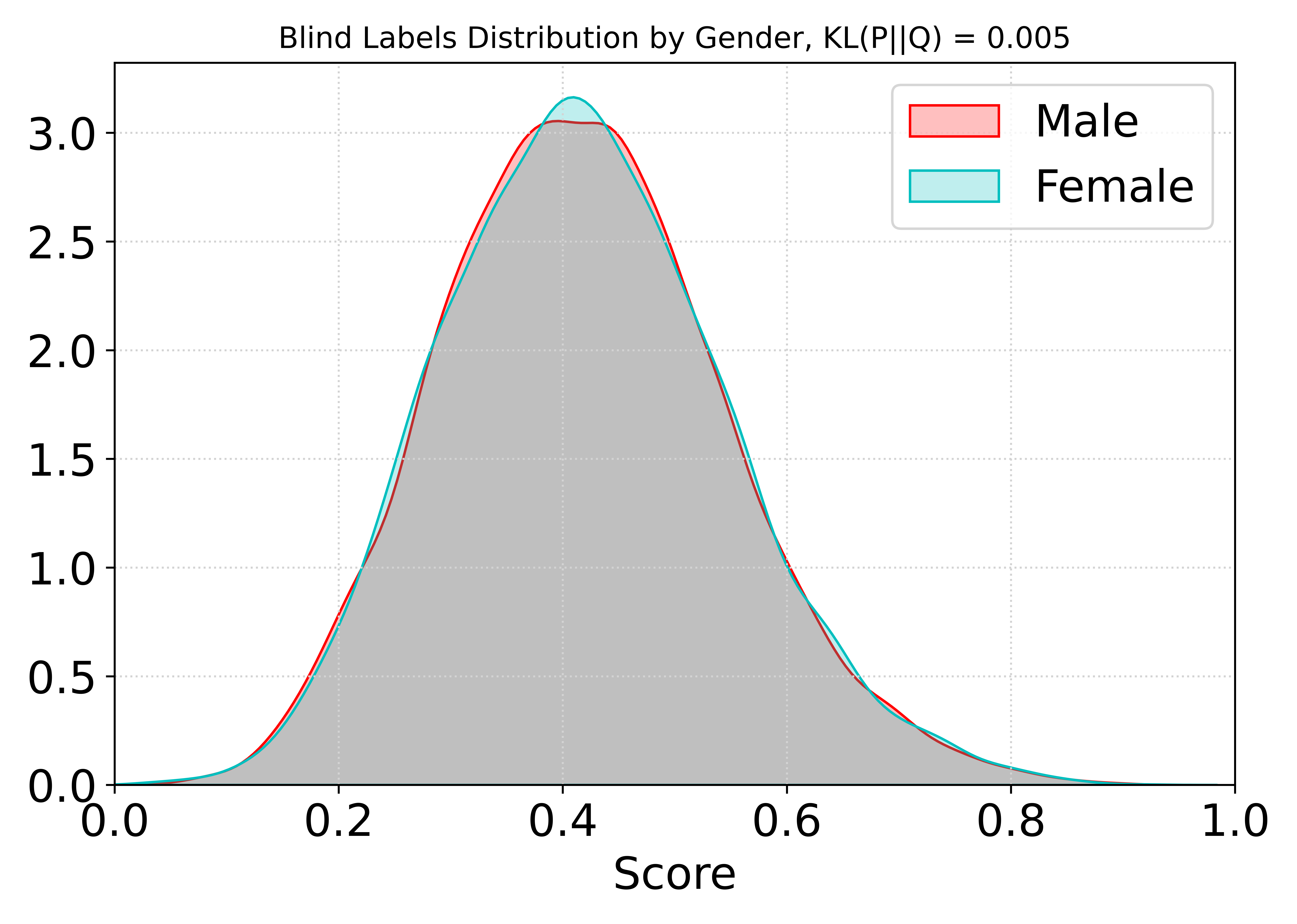}};
                    \node[anchor=center, yshift=37] at (image.center) {\scriptsize KL = 0.005};
                    \node[anchor=center, yshift=25, xshift=-30] at (image.center) {\scriptsize (1)};
                 \end{tikzpicture} 
                 
                 &

                 \begin{tikzpicture}
                    \node[inner sep=0] (image) at (0,0) {\includegraphics[width=0.22\linewidth, trim=5 0 0 25, clip]{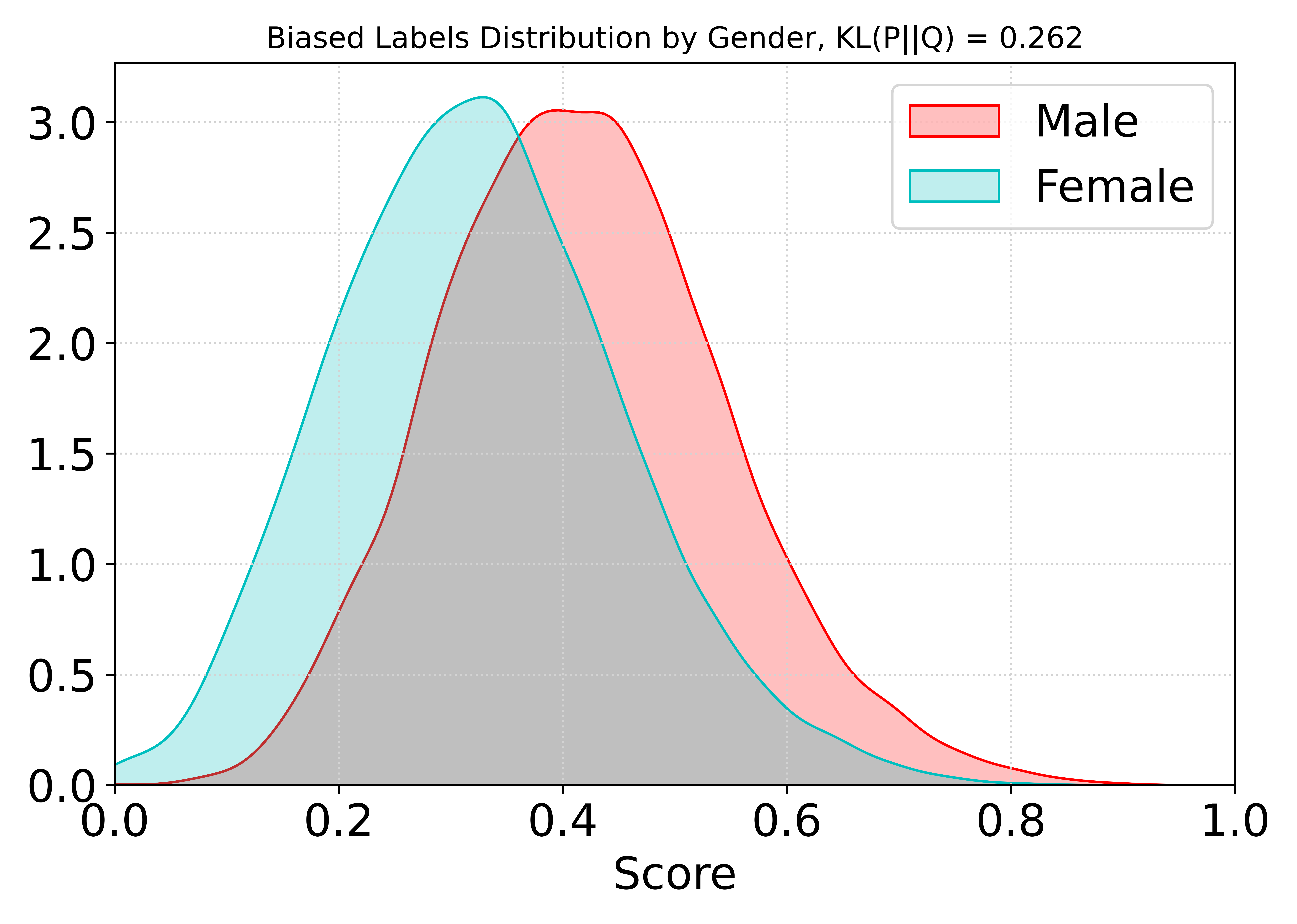}};
                    \node[anchor=center, yshift=37] at (image.center) {\scriptsize KL = 0.262};
                    \node[anchor=center, yshift=25, xshift=-30] at (image.center) {\scriptsize (2)};
                 \end{tikzpicture} 
                
                &

                \begin{tikzpicture}
                    \node[inner sep=0] (image) at (0,0) {\includegraphics[width=0.22\linewidth, trim=5 0 0 25, clip]{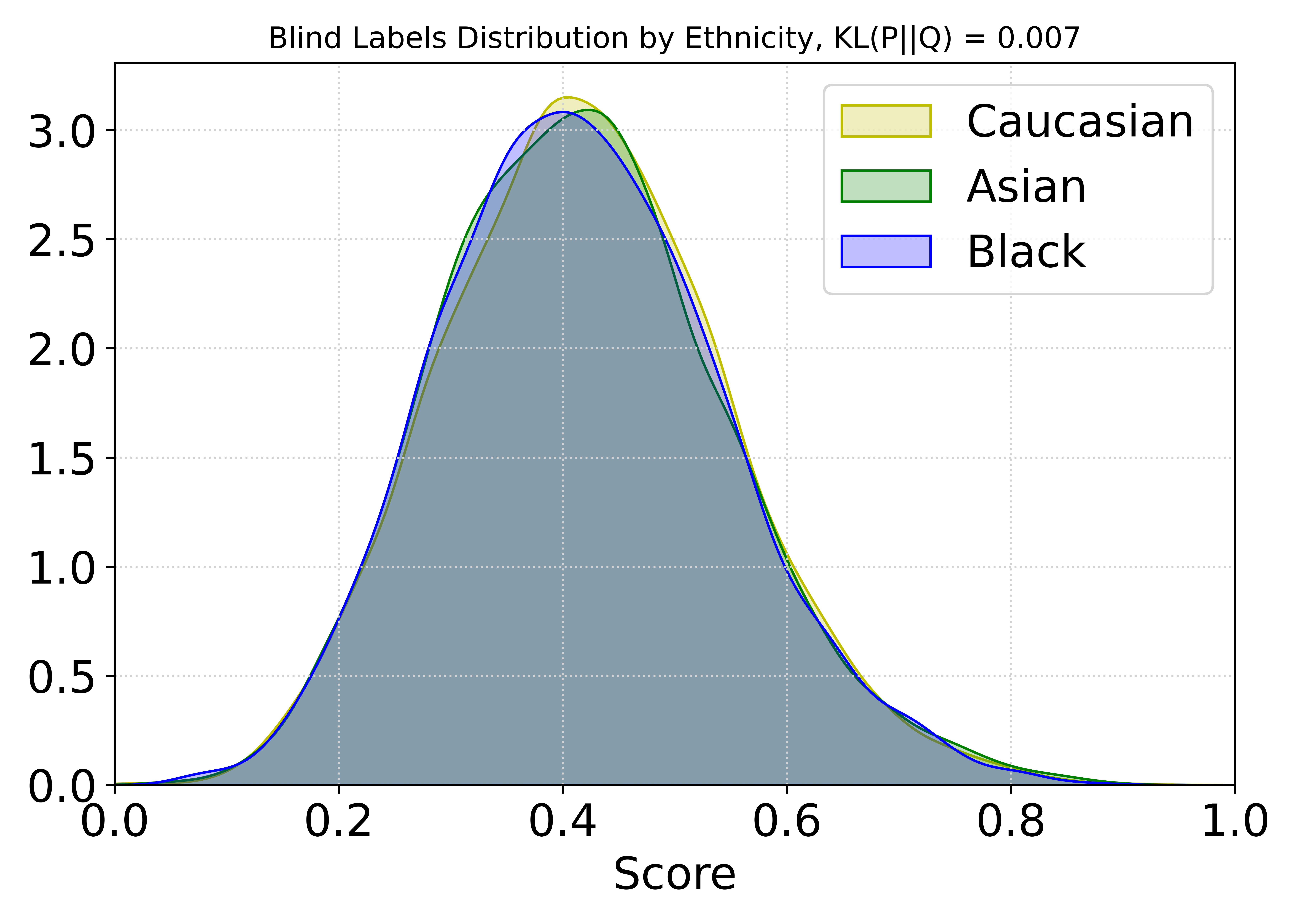}};
                    \node[anchor=center, yshift=37] at (image.center) {\scriptsize KL = 0.007};
                    \node[anchor=center, yshift=25, xshift=-30] at (image.center) {\scriptsize (3)};
                 \end{tikzpicture}
                 
                &

                \begin{tikzpicture}
                    \node[inner sep=0] (image) at (0,0) {\includegraphics[width=0.22\linewidth, trim=5 0 0 25, clip]{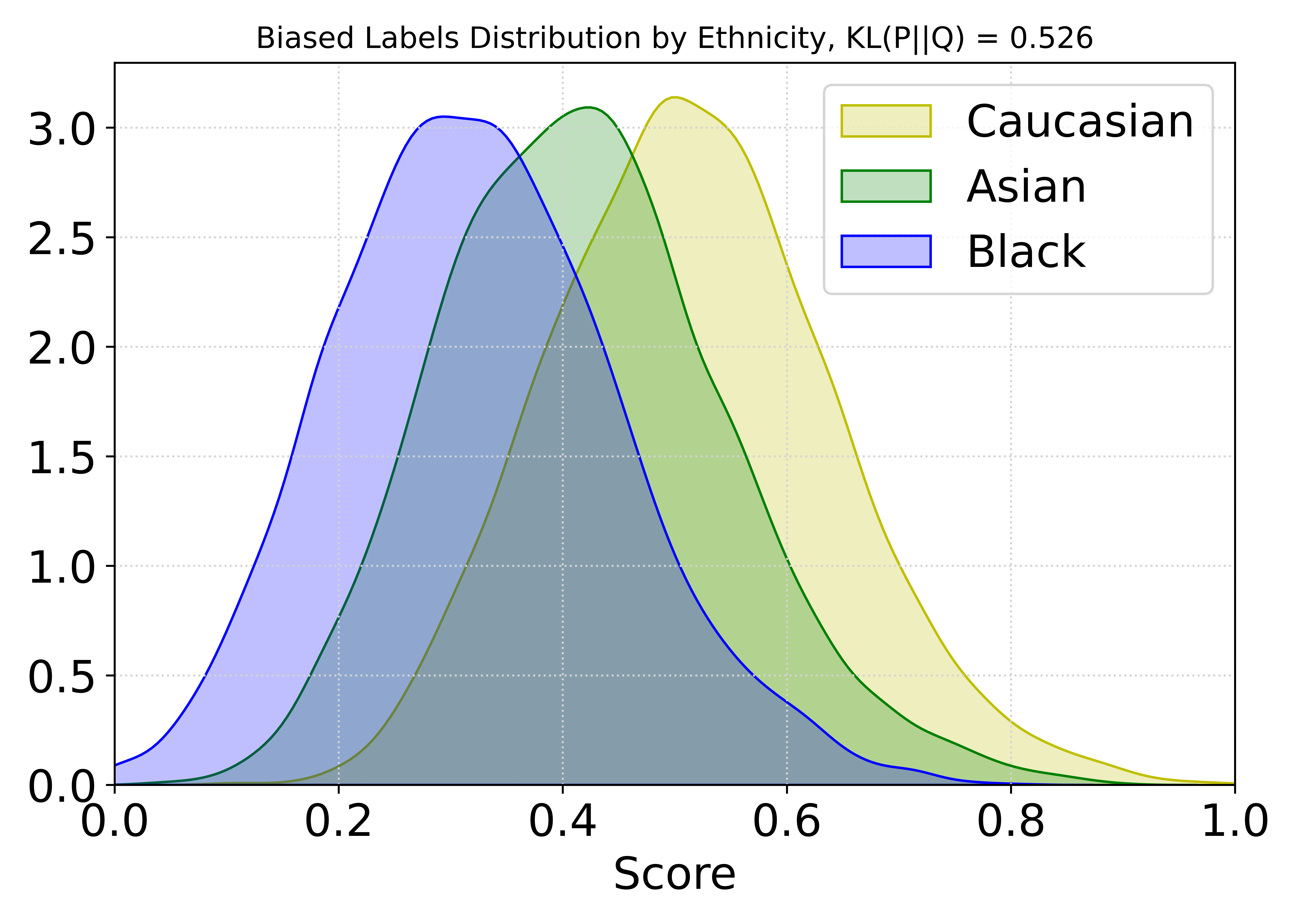}};
                    \node[anchor=center, yshift=37] at (image.center) {\scriptsize KL = 0.526};
                    \node[anchor=center, yshift=25, xshift=-30] at (image.center) {\scriptsize (4)};
                 \end{tikzpicture}
 
                 \vspace{-0.5cm} \\ 

                
                 \rotatebox{90}{\hspace{0.4cm}\textbf{(b)} Tabular}
                 
                 & 
                 
                 \begin{tikzpicture}
                    \node[inner sep=0] (image) at (0,0) {\includegraphics[width=0.22\linewidth, trim=5 0 0 25, clip]{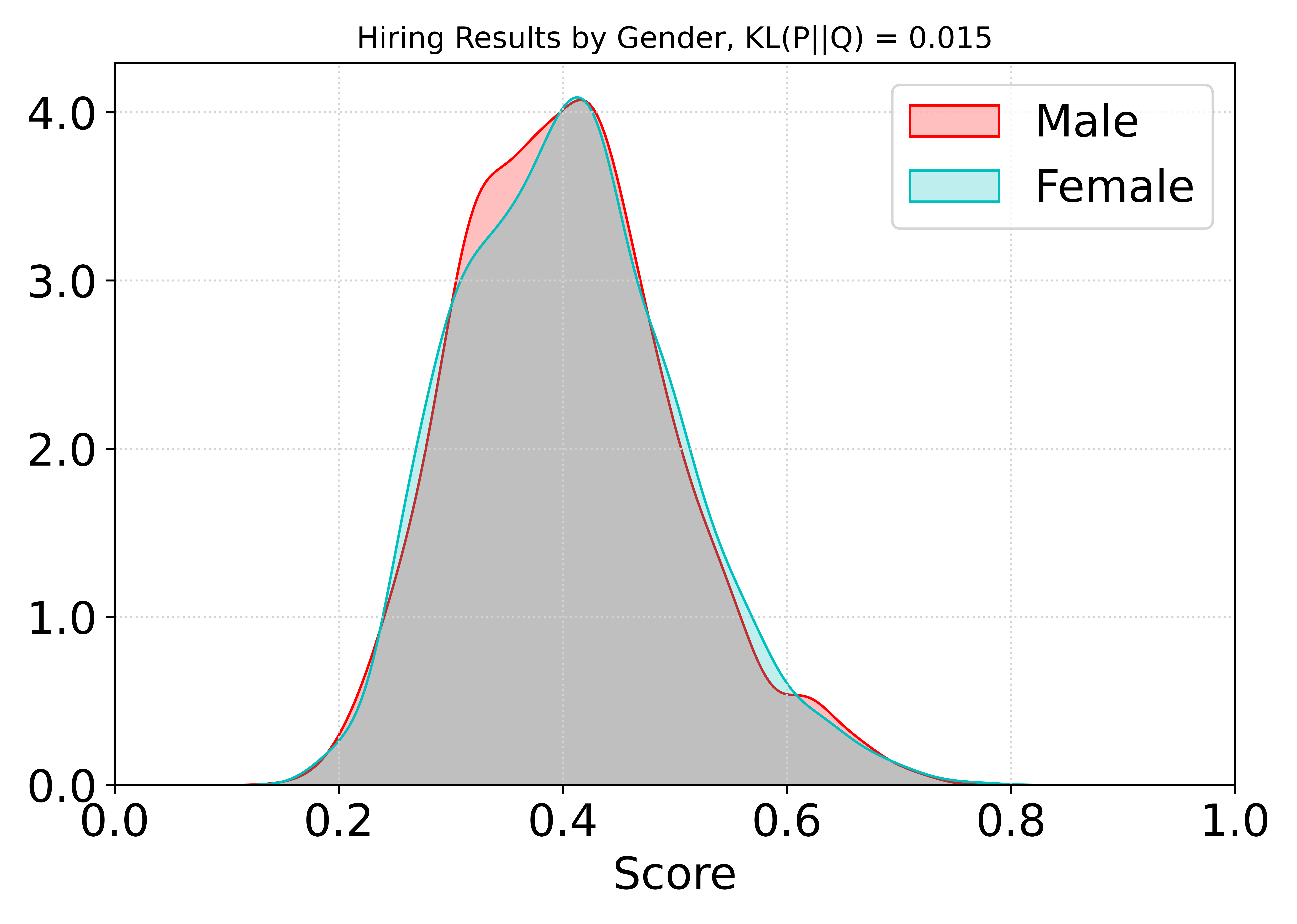}};
                    \node[anchor=center, yshift=37] at (image.center) {\scriptsize KL = 0.015, MAE = 0.060};
                    \node[anchor=center, yshift=25, xshift=-30] at (image.center) {\scriptsize (1)};
                 \end{tikzpicture} 
                 
                 &

                 \begin{tikzpicture}
                    \node[inner sep=0] (image) at (0,0) {\includegraphics[width=0.22\linewidth, trim=5 0 0 25, clip]{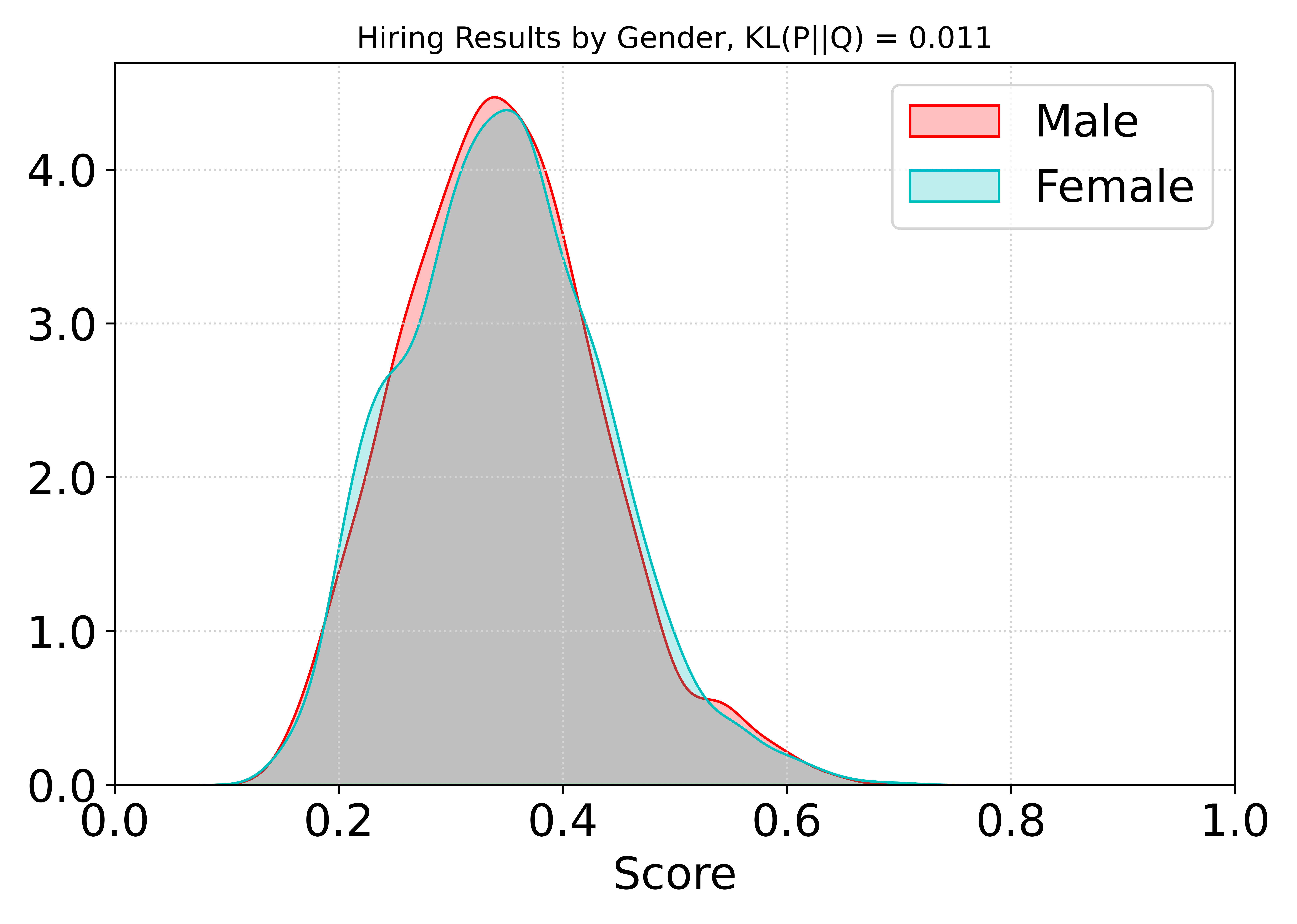}};
                    \node[anchor=center, yshift=37] at (image.center) {\scriptsize \textbf{KL = 0.011}, MAE = 0.073};
                    \node[anchor=center, yshift=25, xshift=-30] at (image.center) {\scriptsize (2)};
                 \end{tikzpicture} 
                
                &

                \begin{tikzpicture}
                    \node[inner sep=0] (image) at (0,0) {\includegraphics[width=0.22\linewidth, trim=5 0 0 25, clip]{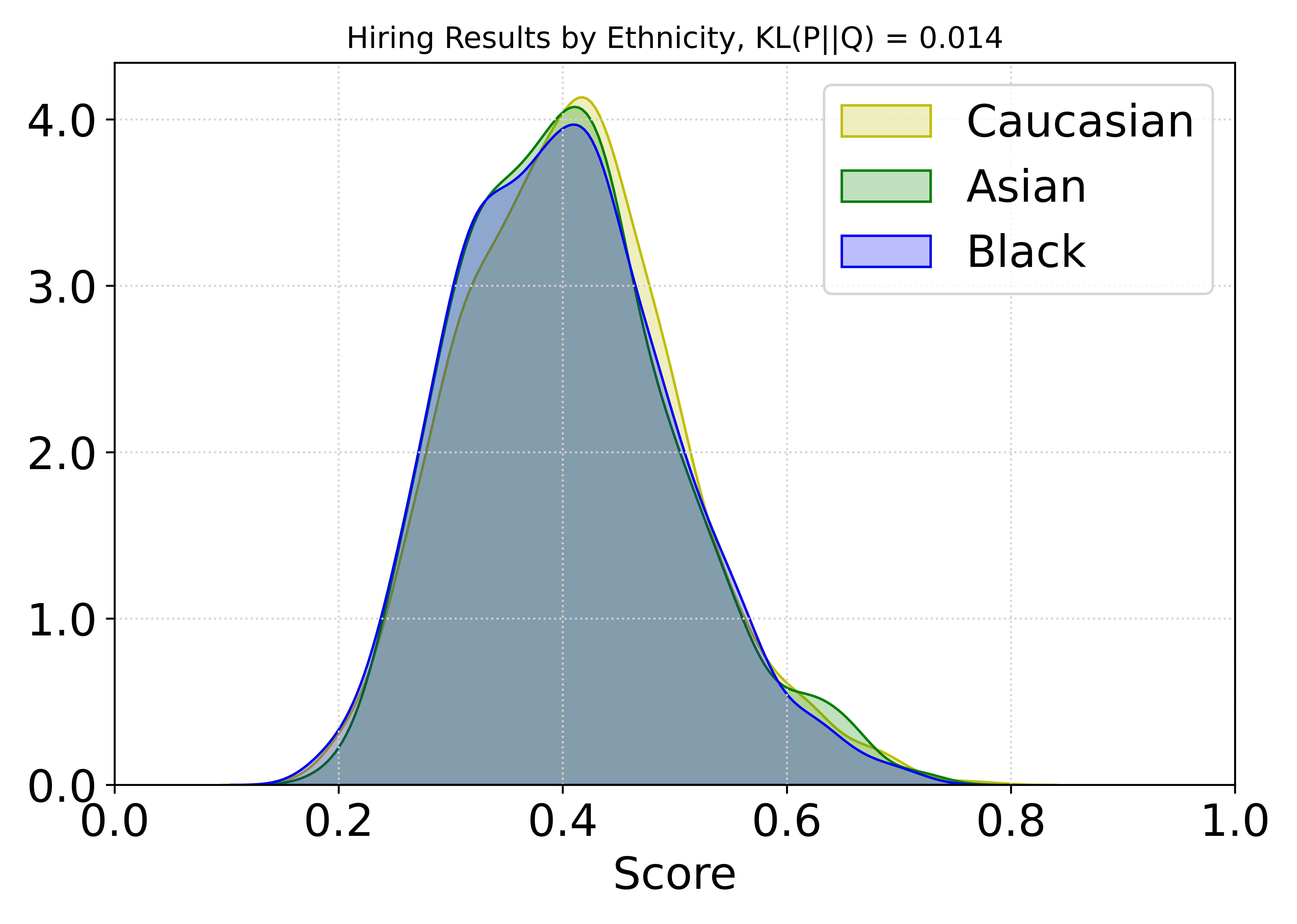}};
                    \node[anchor=center, yshift=37] at (image.center) {\scriptsize KL = 0.014, MAE = 0.060};
                    \node[anchor=center, yshift=25, xshift=-30] at (image.center) {\scriptsize (3)};
                 \end{tikzpicture}
                 
                &

                \begin{tikzpicture}
                    \node[inner sep=0] (image) at (0,0) {\includegraphics[width=0.22\linewidth, trim=5 0 0 25, clip]{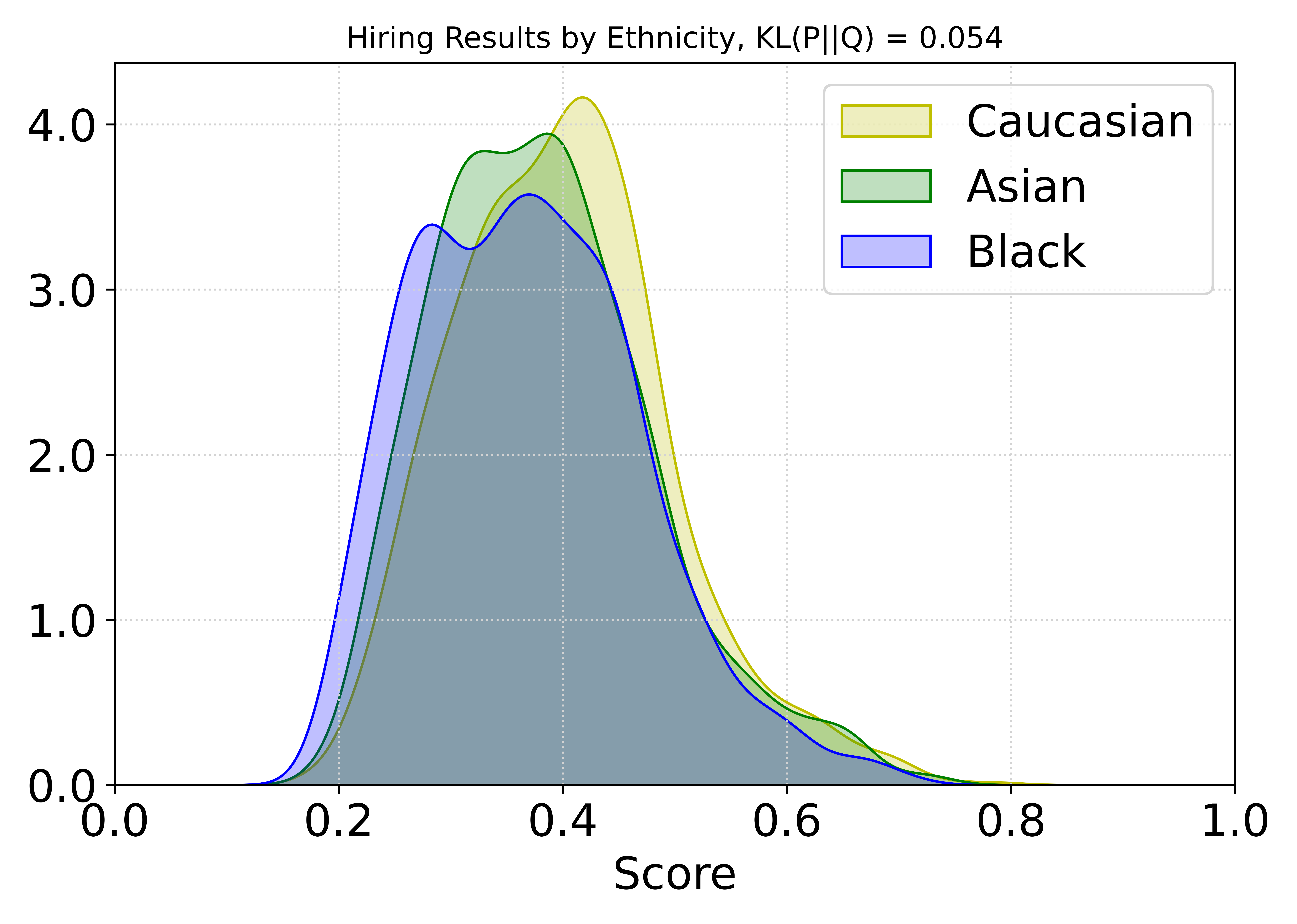}};
                    \node[anchor=center, yshift=37] at (image.center) {\scriptsize KL = 0.048, MAE = 0.082};
                    \node[anchor=center, yshift=25, xshift=-30] at (image.center) {\scriptsize (4)};
                 \end{tikzpicture}
 
                 \vspace{-0.5cm} \\ 


                 \rotatebox{90}{\hspace{0.4cm}\textbf{(c)} Textual}
                 
                 & 
                 
                 \begin{tikzpicture}
                    \node[inner sep=0] (image) at (0,0) {\includegraphics[width=0.22\linewidth, trim=5 0 0 25, clip]{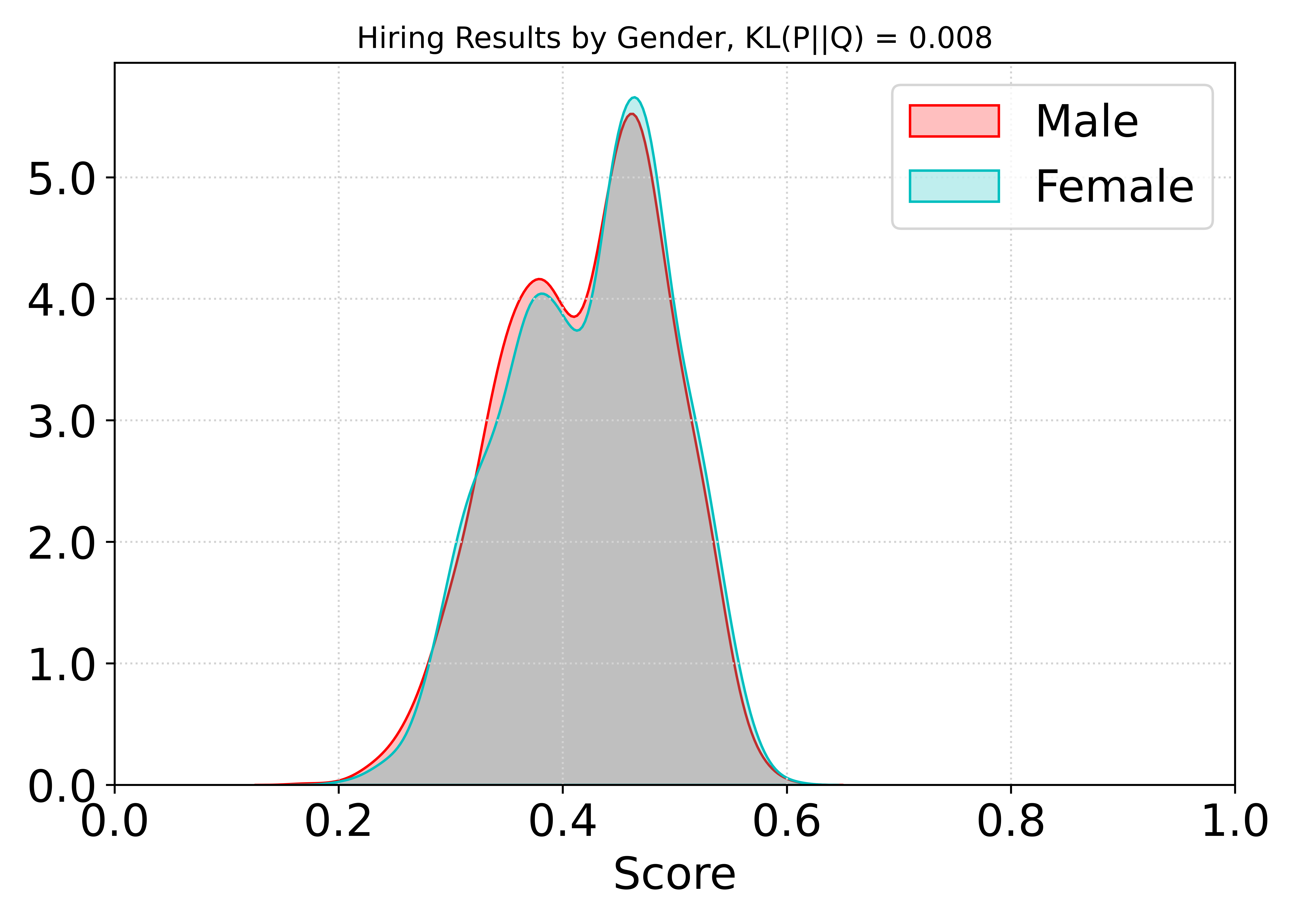}};
                    \node[anchor=center, yshift=37] at (image.center) {\scriptsize KL = 0.008, MAE = 0.092};
                    \node[anchor=center, yshift=25, xshift=-30] at (image.center) {\scriptsize (1)};
                 \end{tikzpicture} 
                 
                 &

                 \begin{tikzpicture}
                    \node[inner sep=0] (image) at (0,0) {\includegraphics[width=0.22\linewidth, trim=5 0 0 25, clip]{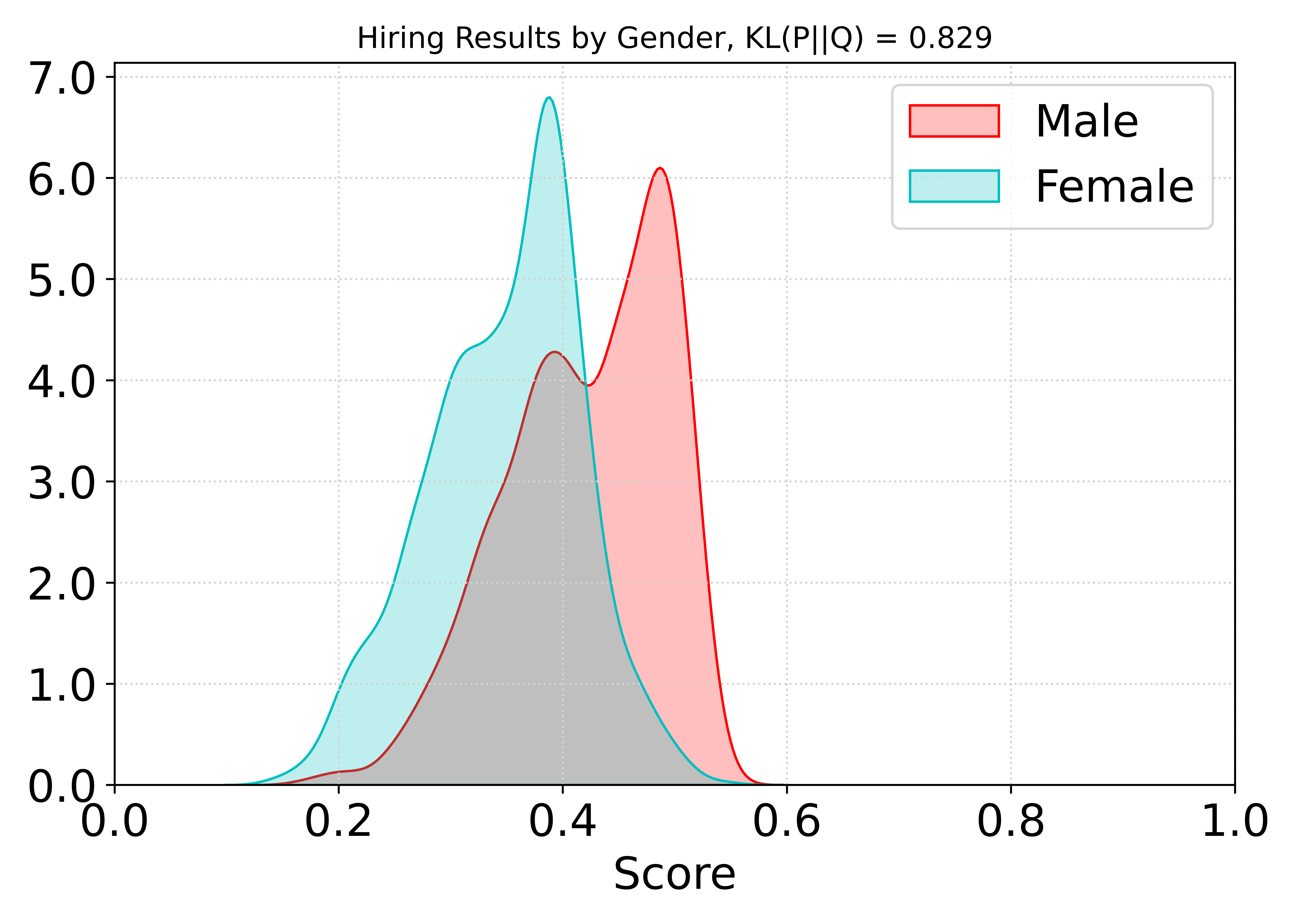}};
                    \node[anchor=center, yshift=37] at (image.center) {\scriptsize KL = 0.829, MAE = 0.079};
                    \node[anchor=center, yshift=25, xshift=-30] at (image.center) {\scriptsize (2)};
                 \end{tikzpicture} 
                
                &

                \begin{tikzpicture}
                    \node[inner sep=0] (image) at (0,0) {\includegraphics[width=0.22\linewidth, trim=5 0 0 25, clip]{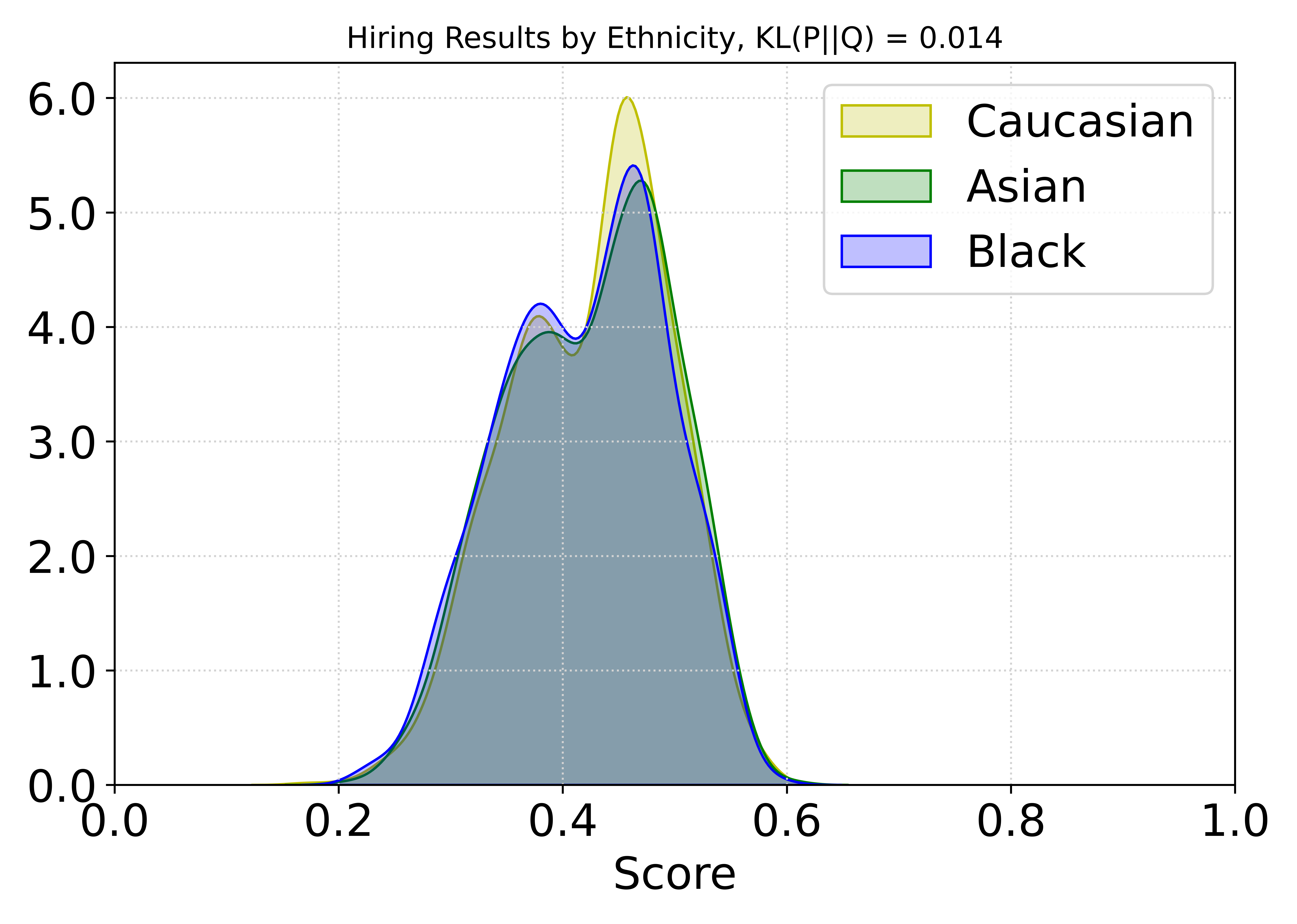}};
                    \node[anchor=center, yshift=37] at (image.center) {\scriptsize KL = 0.014, MAE = 0.092};
                    \node[anchor=center, yshift=25, xshift=-30] at (image.center) {\scriptsize (3)};
                 \end{tikzpicture}
                 
                &

                \begin{tikzpicture}
                    \node[inner sep=0] (image) at (0,0) {\includegraphics[width=0.22\linewidth, trim=5 0 0 25, clip]{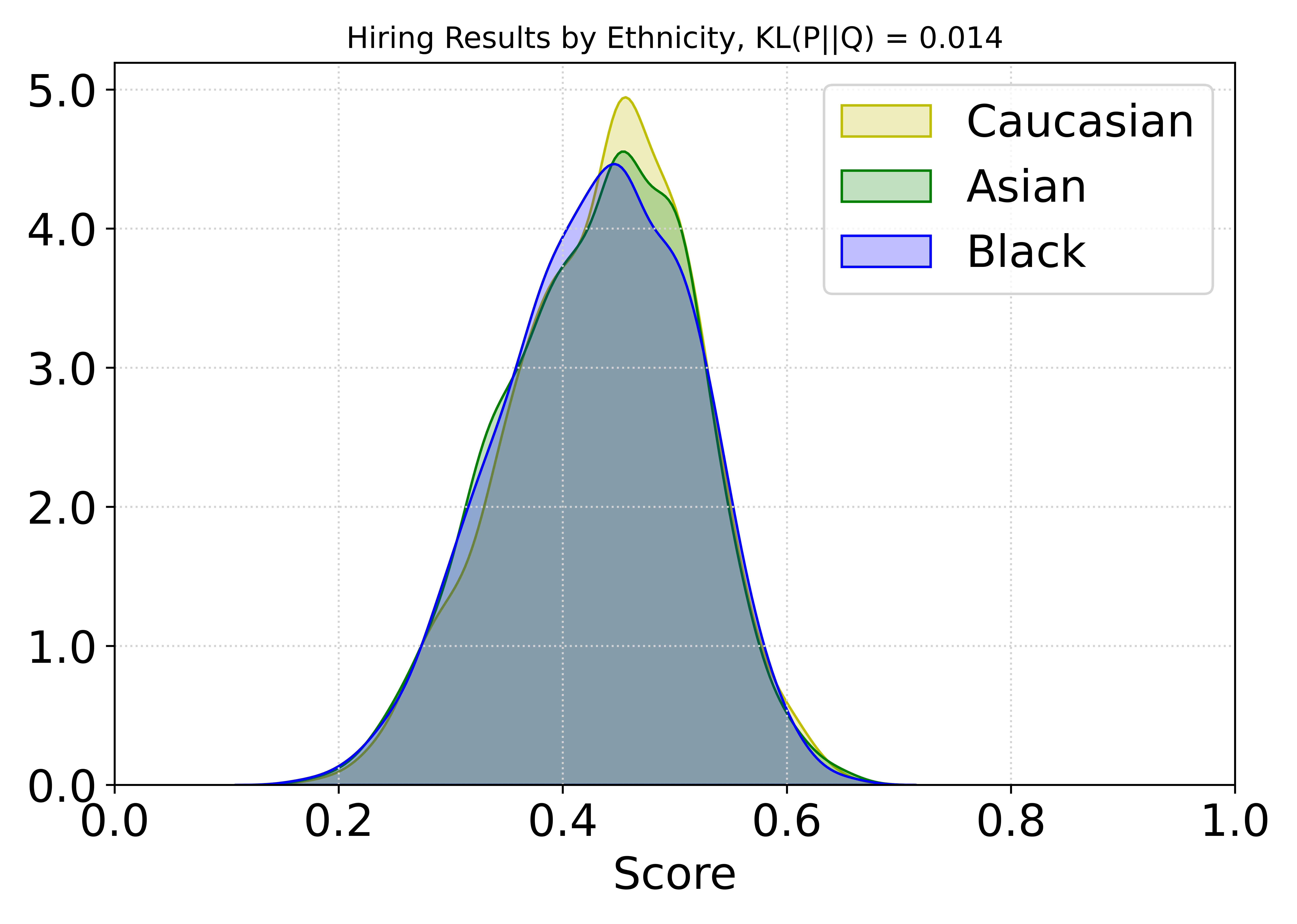}};
                    \node[anchor=center, yshift=37] at (image.center) {\scriptsize \textbf{KL = 0.017}, MAE = 0.115};
                    \node[anchor=center, yshift=25, xshift=-30] at (image.center) {\scriptsize (4)};
                 \end{tikzpicture}
 
                 \vspace{-0.5cm} \\ 


                 \rotatebox{90}{\hspace{0.5cm}\textbf{(d)} Visual}
                 
                 & 
                 
                 \begin{tikzpicture}
                    \node[inner sep=0] (image) at (0,0) {\includegraphics[width=0.22\linewidth, trim=5 0 0 20, clip]{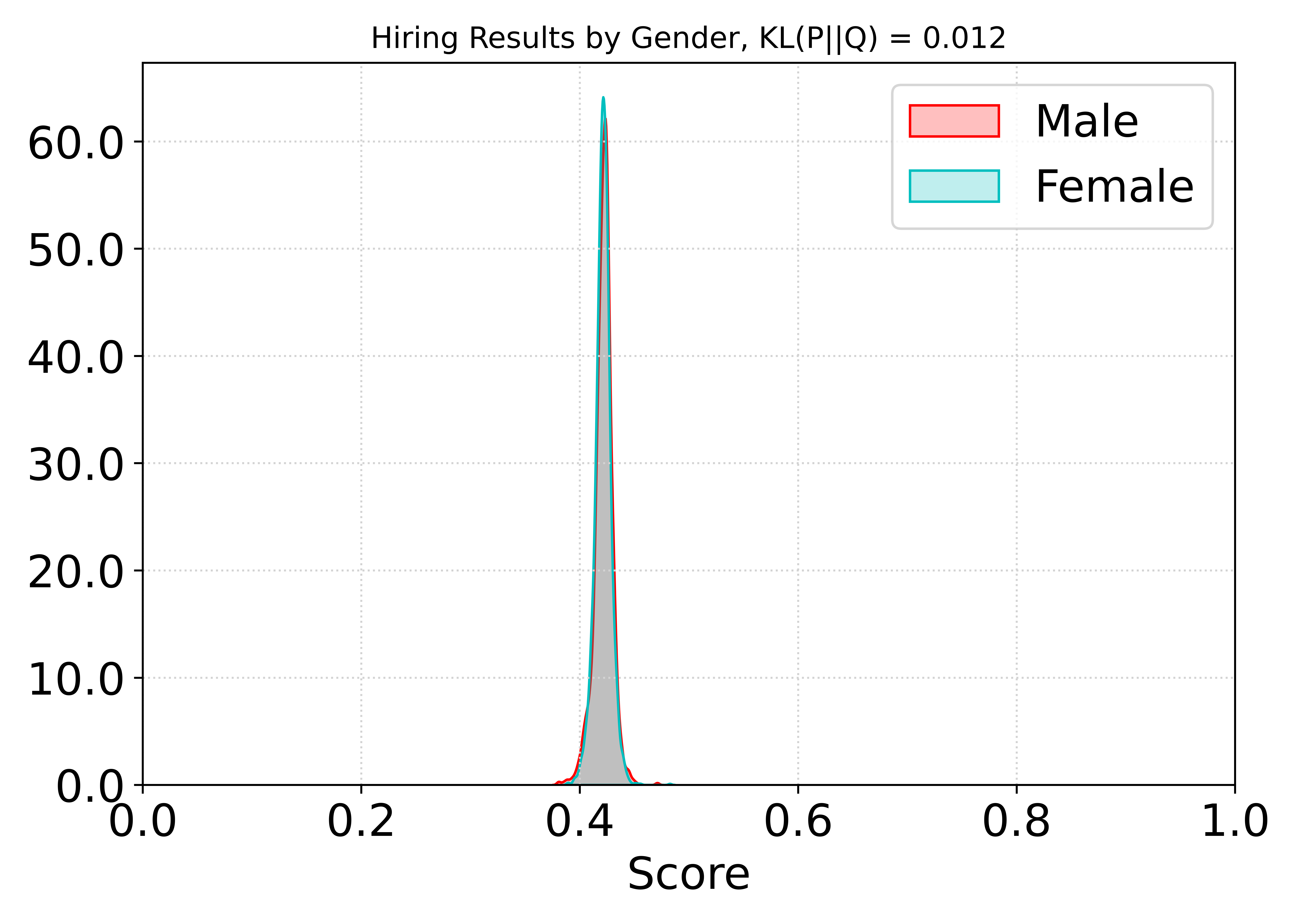}};
                    \node[anchor=center, yshift=37] at (image.center) {\scriptsize KL = 0.012, MAE = 0.100};
                    \node[anchor=center, yshift=25, xshift=-30] at (image.center) {\scriptsize (1)};
                 \end{tikzpicture} 
                 
                 &

                 \begin{tikzpicture}
                    \node[inner sep=0] (image) at (0,0) {\includegraphics[width=0.22\linewidth, trim=5 0 0 20, clip]{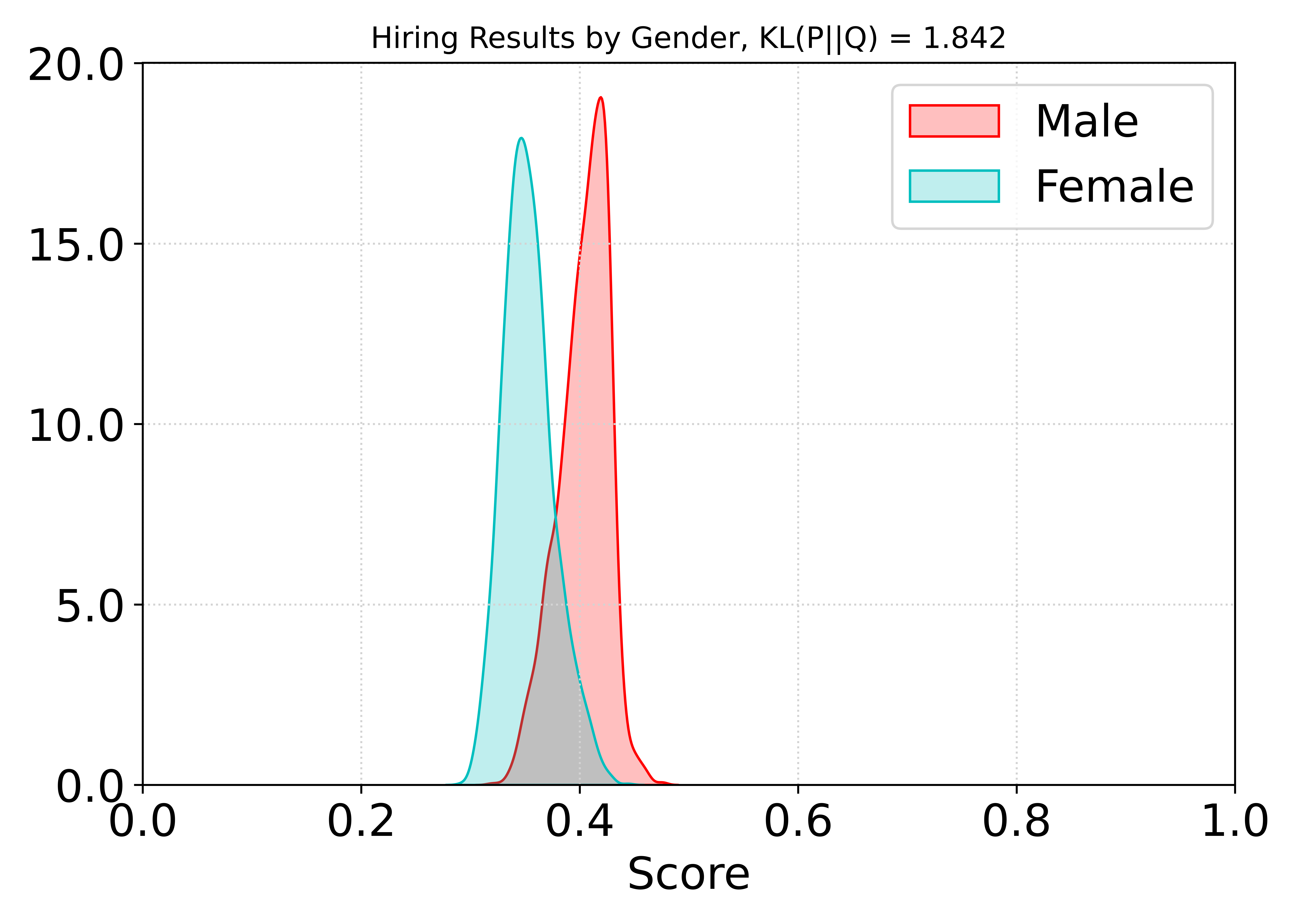}};
                    \node[anchor=center, yshift=37] at (image.center) {\scriptsize KL = 1.842, MAE = 0.104};
                    \node[anchor=center, yshift=25, xshift=-30] at (image.center) {\scriptsize (2)};
                 \end{tikzpicture} 
                
                &

                \begin{tikzpicture}
                    \node[inner sep=0] (image) at (0,0) {\includegraphics[width=0.22\linewidth, trim=5 0 0 20, clip]{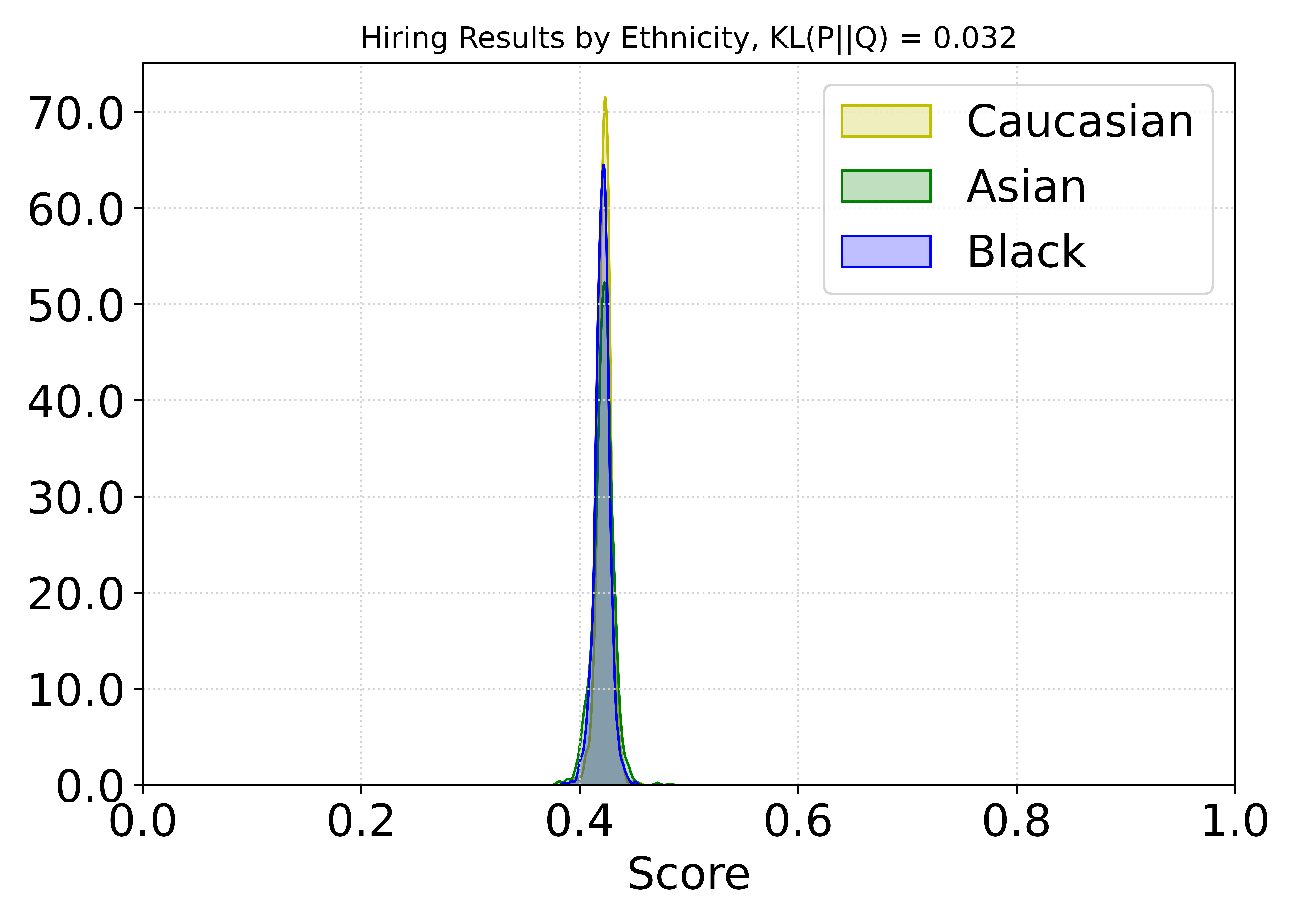}};
                    \node[anchor=center, yshift=37] at (image.center) {\scriptsize KL = 0.032, MAE = 0.100};
                    \node[anchor=center, yshift=25, xshift=-30] at (image.center) {\scriptsize (3)};
                 \end{tikzpicture}
                 
                &

                \begin{tikzpicture}
                    \node[inner sep=0] (image) at (0,0) {\includegraphics[width=0.22\linewidth, trim=5 0 0 20, clip]{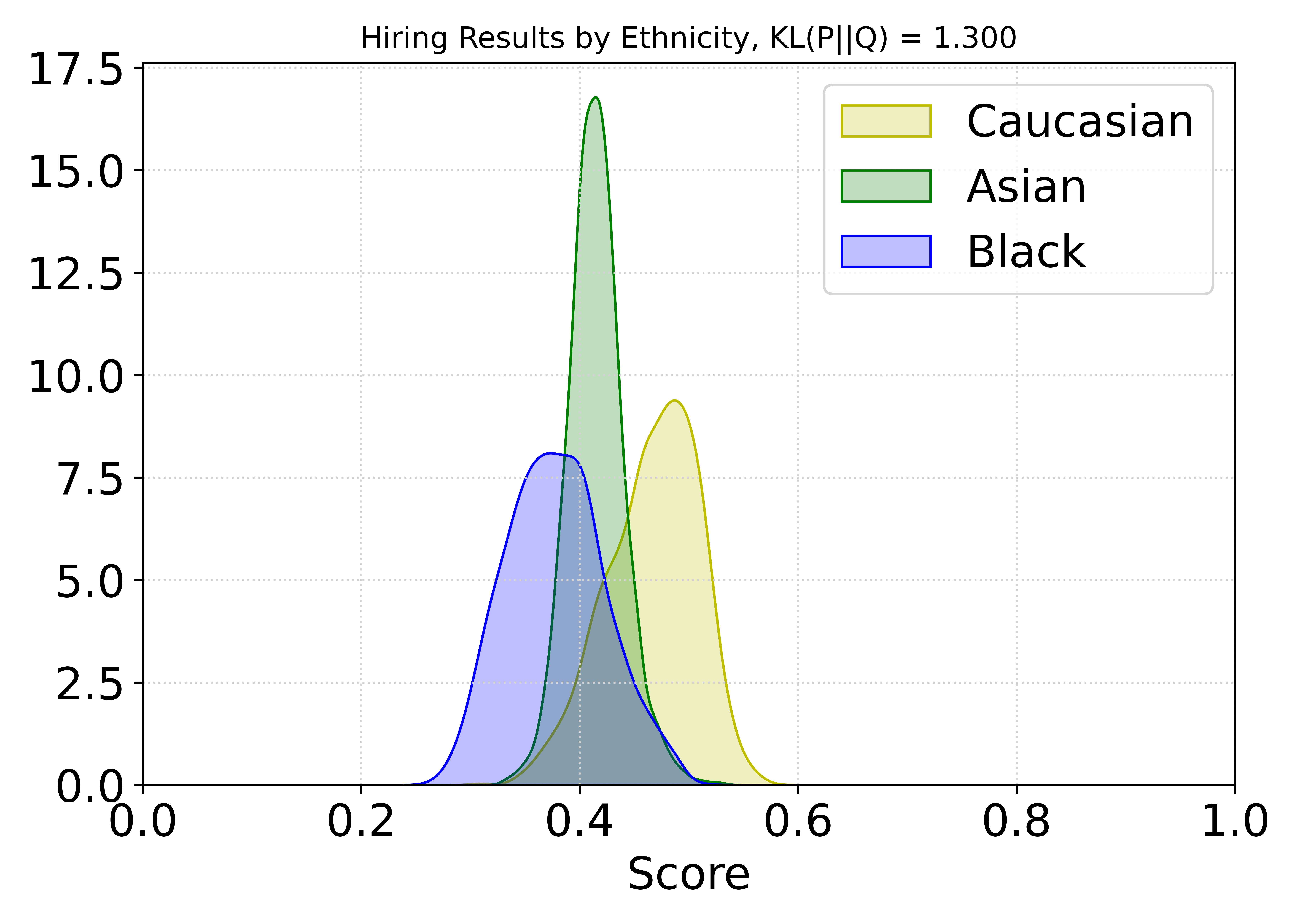}};
                    \node[anchor=center, yshift=37] at (image.center) {\scriptsize KL = 0.973, MAE = 0.111};
                    \node[anchor=center, yshift=25, xshift=-30] at (image.center) {\scriptsize (4)};
                 \end{tikzpicture}
 
                 \vspace{-0.5cm} \\ 


                 \rotatebox{90}{\hspace{0cm}\textbf{(e)} Early-Fusion}
                 
                 & 
                 
                 \begin{tikzpicture}
                    \node[inner sep=0] (image) at (0,0) {\includegraphics[width=0.22\linewidth, trim=5 0 0 25, clip]{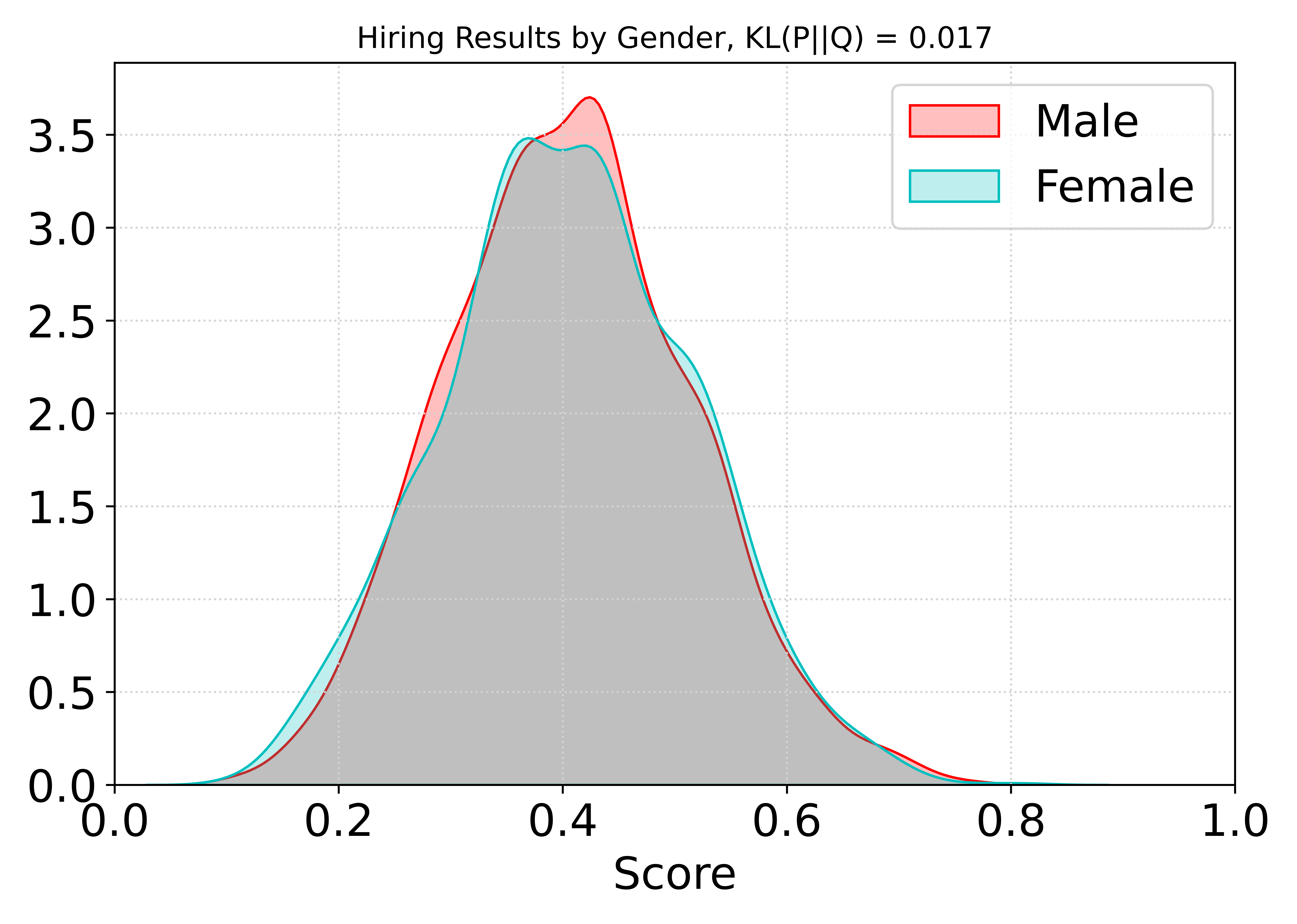}};
                    \node[anchor=center, yshift=37] at (image.center) {\scriptsize KL = 0.017, \textbf{MAE = 0.026}};
                    \node[anchor=center, yshift=25, xshift=-30] at (image.center) {\scriptsize (1)};
                 \end{tikzpicture} 
                 
                 &

                 \begin{tikzpicture}
                    \node[inner sep=0] (image) at (0,0) {\includegraphics[width=0.22\linewidth, trim=5 0 0 25, clip]{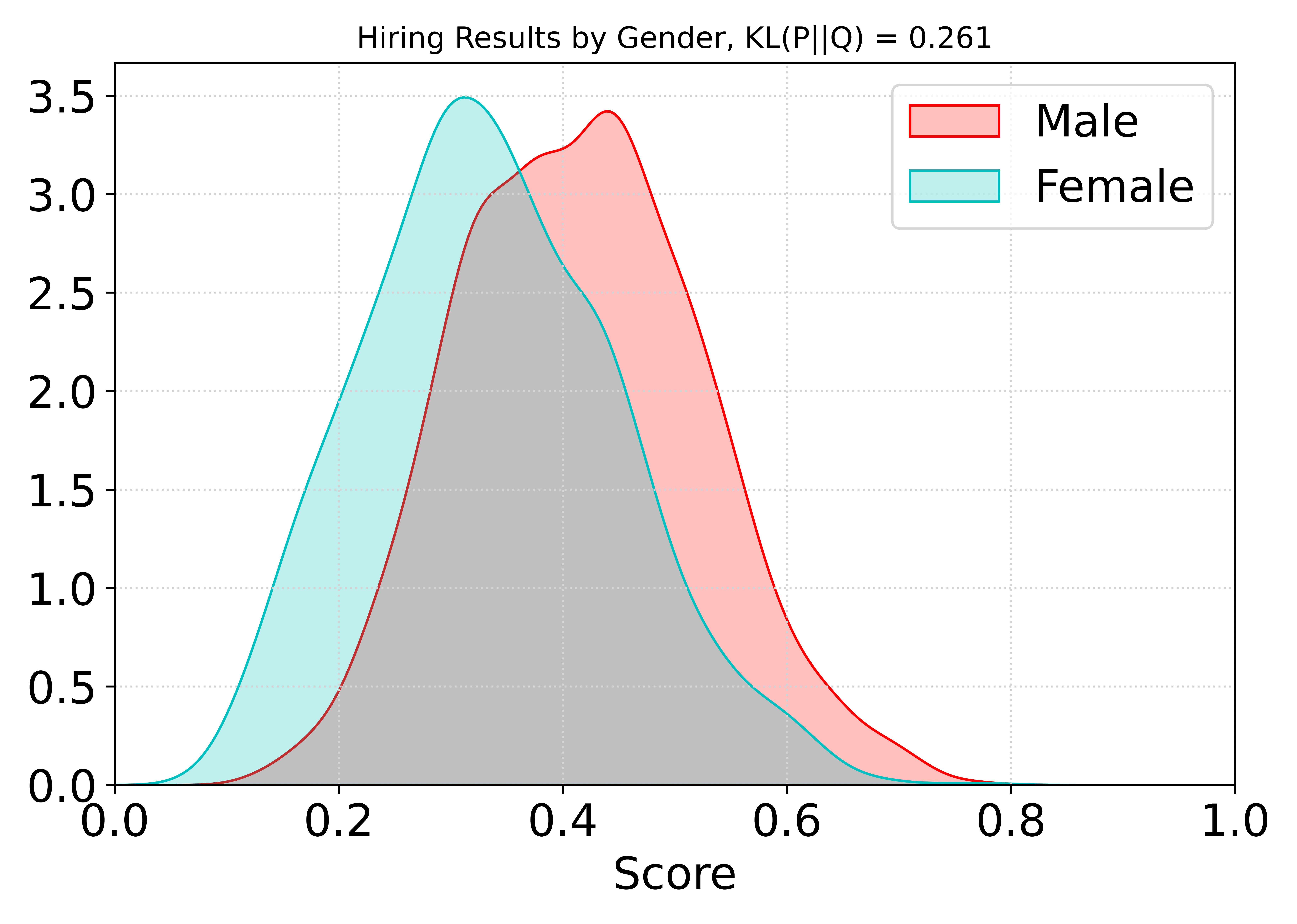}};
                    \node[anchor=center, yshift=37] at (image.center) {\scriptsize KL = 0.261, \textbf{MAE = 0.032}};
                    \node[anchor=center, yshift=25, xshift=-30] at (image.center) {\scriptsize (2)};
                 \end{tikzpicture} 
                
                &

                \begin{tikzpicture}
                    \node[inner sep=0] (image) at (0,0) {\includegraphics[width=0.22\linewidth, trim=5 0 0 25, clip]{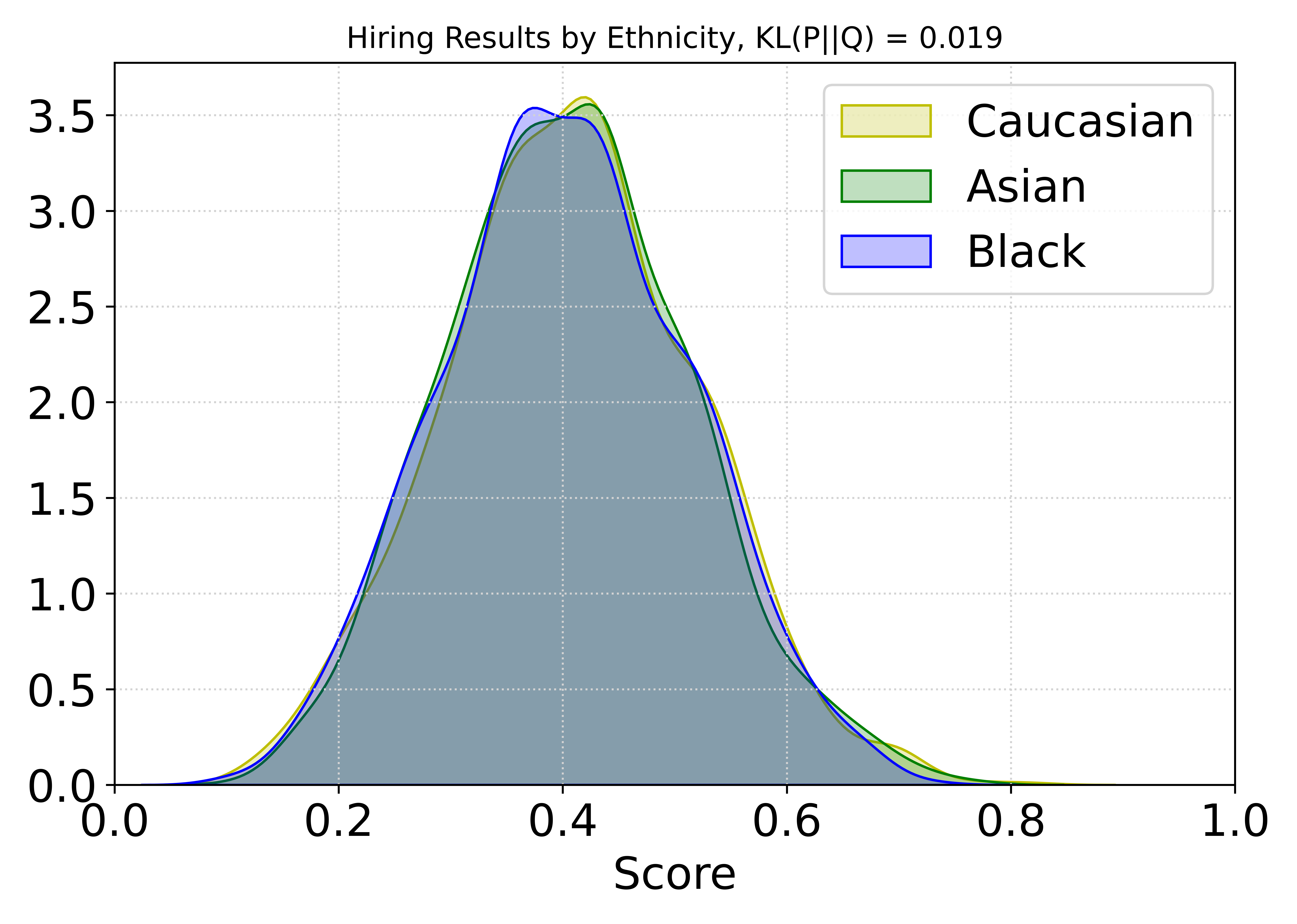}};
                    \node[anchor=center, yshift=37] at (image.center) {\scriptsize KL = 0.019, \textbf{MAE = 0.026}};
                    \node[anchor=center, yshift=25, xshift=-30] at (image.center) {\scriptsize (3)};
                 \end{tikzpicture}
                 
                &

                \begin{tikzpicture}
                    \node[inner sep=0] (image) at (0,0) {\includegraphics[width=0.22\linewidth, trim=5 0 0 25, clip]{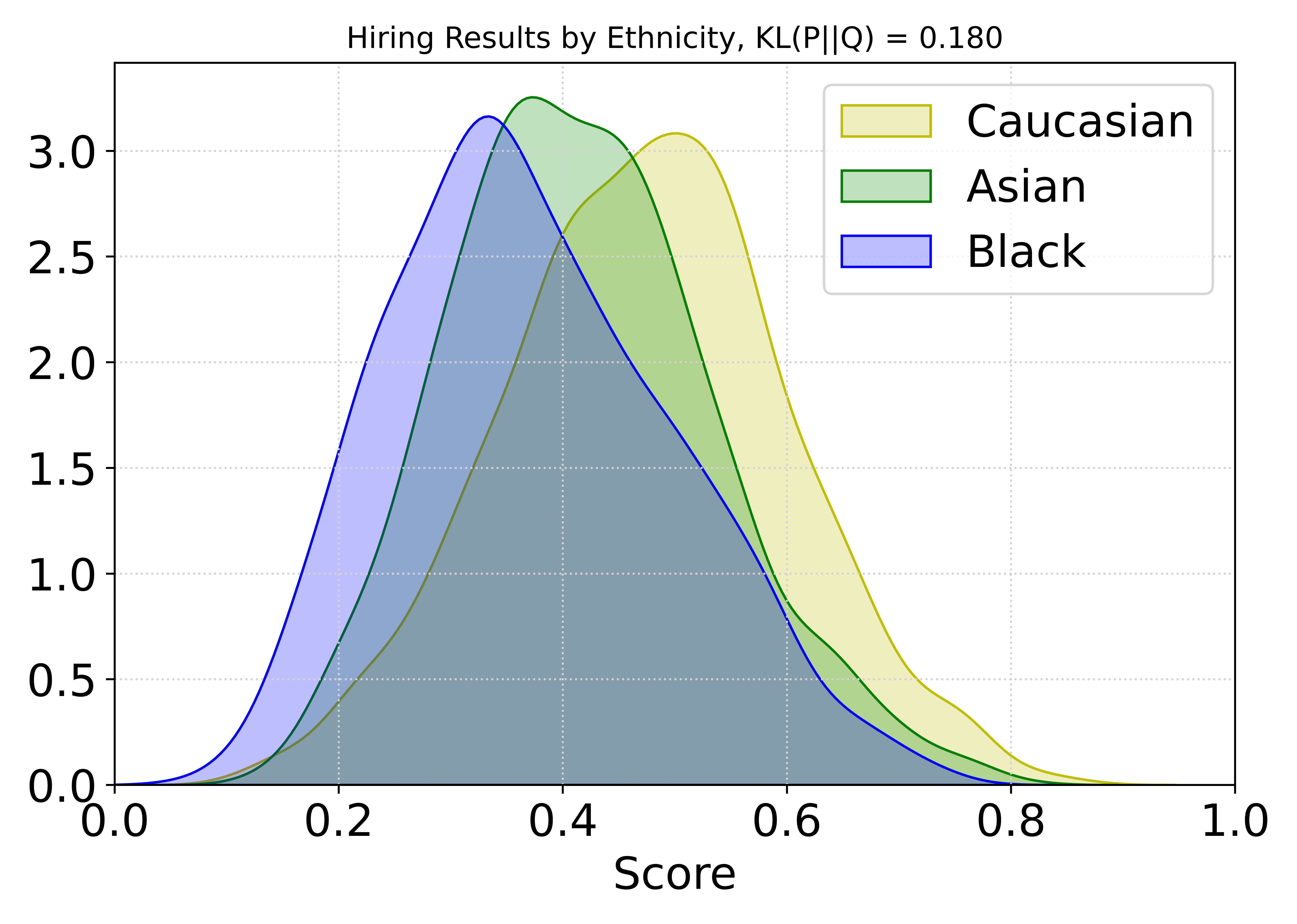}};
                    \node[anchor=center, yshift=37] at (image.center) {\scriptsize KL = 0.174, \textbf{MAE = 0.059}};
                    \node[anchor=center, yshift=25, xshift=-30] at (image.center) {\scriptsize (4)};
                 \end{tikzpicture}
 
                 \vspace{-0.5cm} \\ 


                 \rotatebox{90}{\hspace{0.1cm}\textbf{(f)} Late-Fusion}
                 
                 & 
                 
                 \begin{tikzpicture}
                    \node[inner sep=0] (image) at (0,0) {\includegraphics[width=0.22\linewidth, trim=5 0 0 19, clip]{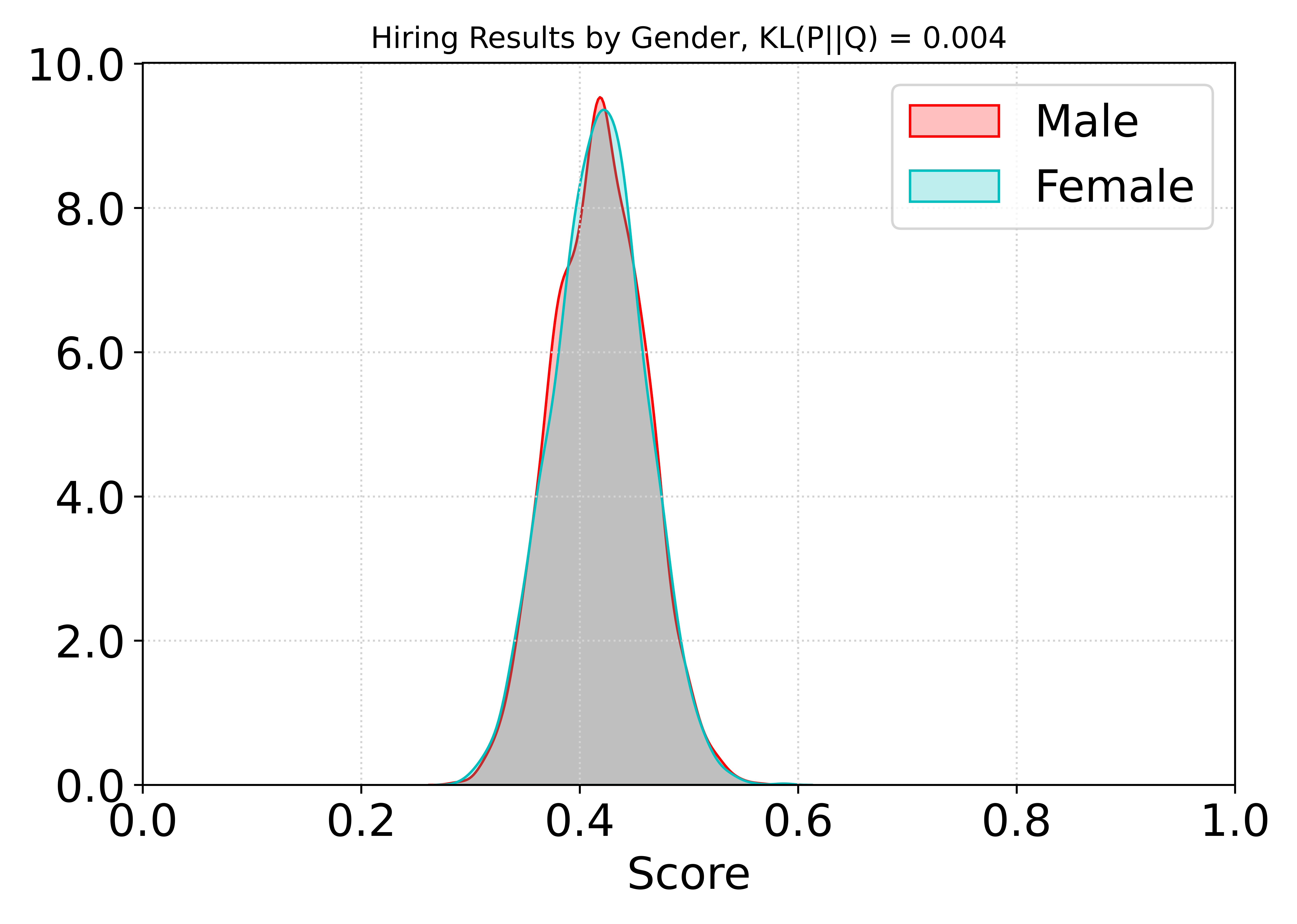}};
                    \node[anchor=center, yshift=37] at (image.center) {\scriptsize \textbf{KL = 0.004}, MAE = 0.084};
                    \node[anchor=center, yshift=25, xshift=-30] at (image.center) {\scriptsize (1)};
                 \end{tikzpicture} 
                 
                 &

                 \begin{tikzpicture}
                    \node[inner sep=0] (image) at (0,0) {\includegraphics[width=0.22\linewidth, trim=5 0 0 19, clip]{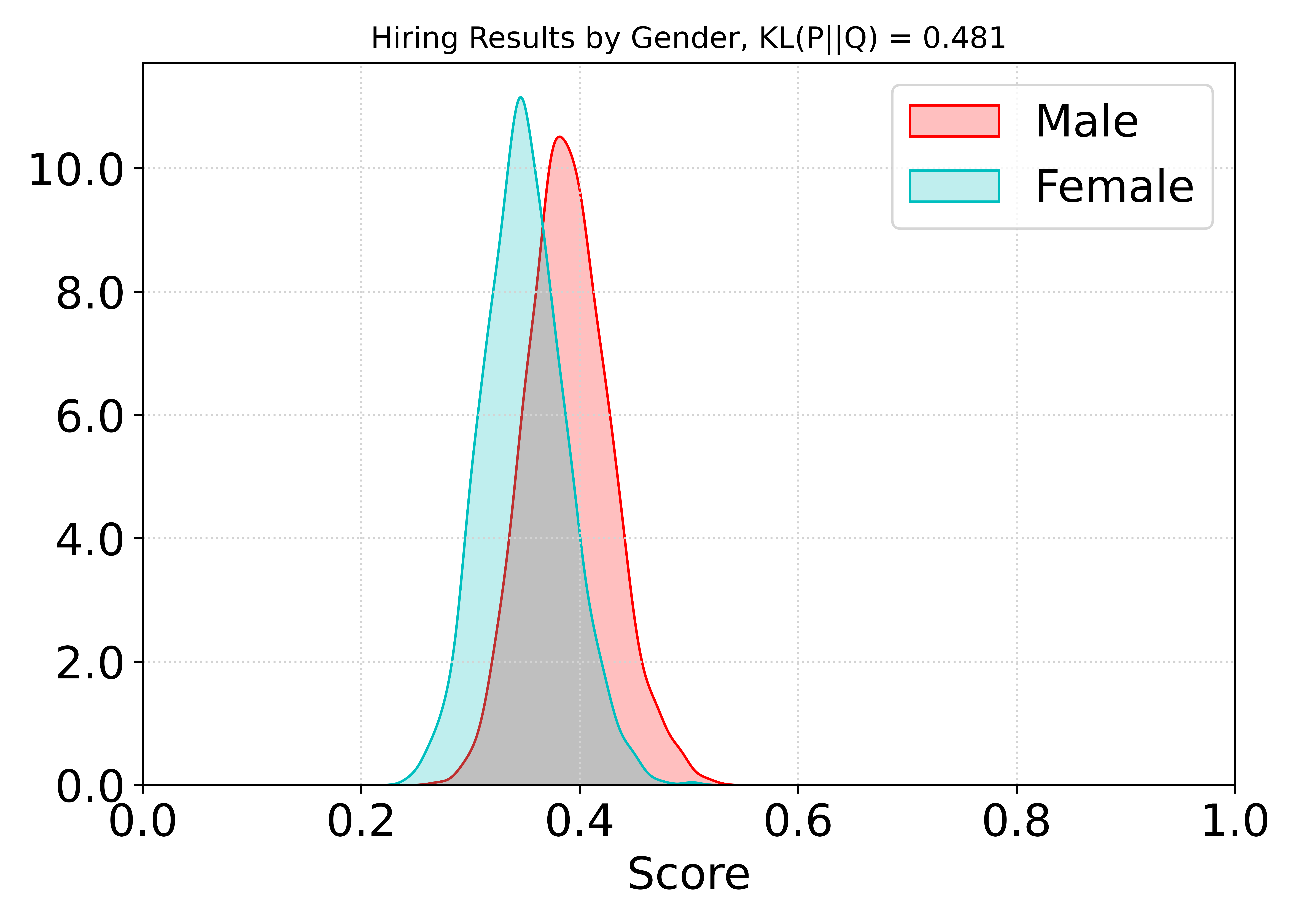}};
                    \node[anchor=center, yshift=37] at (image.center) {\scriptsize KL = 0.481, MAE = 0.090};
                    \node[anchor=center, yshift=25, xshift=-30] at (image.center) {\scriptsize (2)};
                 \end{tikzpicture} 
                
                &

                \begin{tikzpicture}
                    \node[inner sep=0] (image) at (0,0) {\includegraphics[width=0.22\linewidth, trim=5 0 0 19, clip]{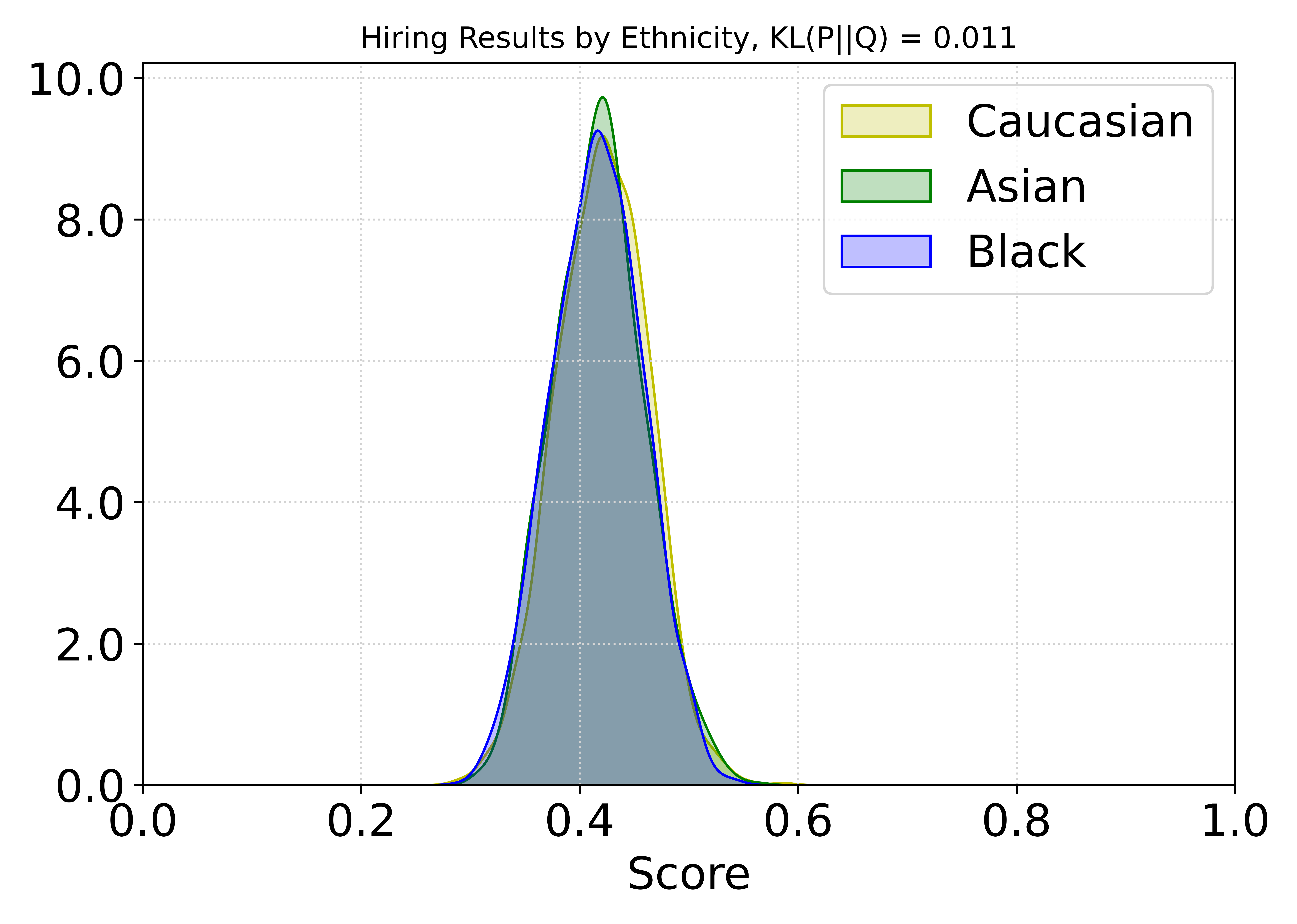}};
                    \node[anchor=center, yshift=37] at (image.center) {\scriptsize \textbf{KL = 0.011}, MAE = 0.084};
                    \node[anchor=center, yshift=25, xshift=-30] at (image.center) {\scriptsize (3)};
                 \end{tikzpicture}
                 
                &

                \begin{tikzpicture}
                    \node[inner sep=0] (image) at (0,0) {\includegraphics[width=0.22\linewidth, trim=5 0 0 19, clip]{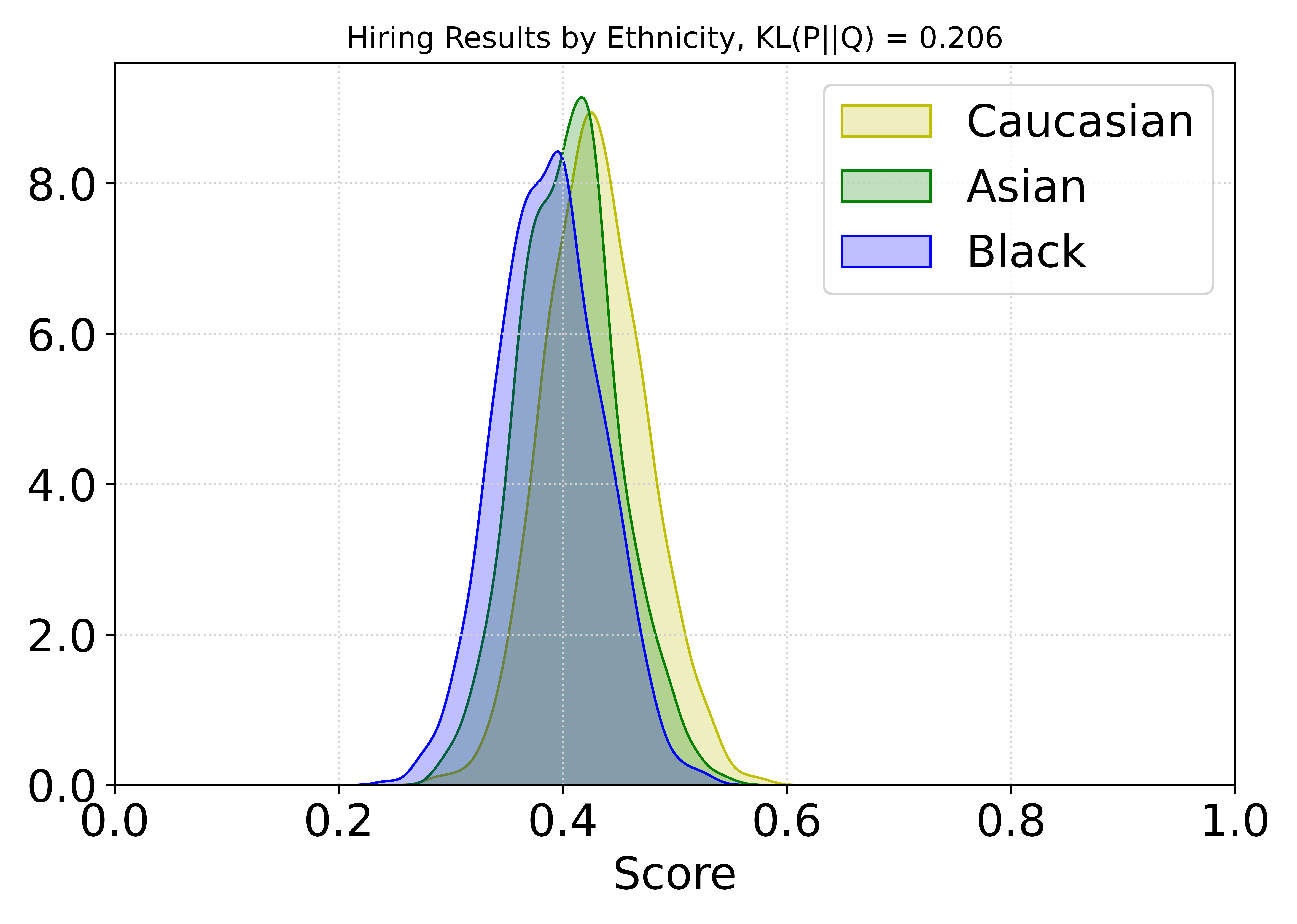}};
                    \node[anchor=center, yshift=37] at (image.center) {\scriptsize KL = 0.206, MAE = 0.103};
                    \node[anchor=center, yshift=25, xshift=-30] at (image.center) {\scriptsize (4)};
                 \end{tikzpicture}
 
                 \vspace{-0.3cm} \\ 

        
                {} & \textcolor{Black}{Neutral} & \textcolor{Maroon}{Gender-Biased} & \textcolor{Black}{Neutral} & \textcolor{DarkBlue}{Ethnicity-Biased} \\
                
            \end{tabular}
            \caption{KL-divergence between score distributions across Gender and Ethnicity demographics for different modalities and bias setups. Lower KL and MAE scores are better.}
            
            \label{fig:KL-divergence}
        \end{figure}
     \end{center} \vspace{-0.7cm} 

\noindent \textbf{In unbiased ideal-world (Neutral):} We note that the \emph{ground-truth} distributions are closely aligned for both demographics (c.f., Figure~\ref{fig:KL-divergence}a). W.r.t. individual modalities (c.f., Figure~\ref{fig:KL-divergence}b - ~\ref{fig:KL-divergence}d), we can see that \emph{tabular} modality exhibits a lower score distribution centred around a mean of $0.4$ with a negatively-skewed distribution, indicating that it tends to underestimate the \emph{ground-truth}. The presence of a bimodal distribution in the \emph{textual} modality is specially intriguing, demonstrating its ability to differentiate between instances with high and low scores. The \emph{Visual} modality, on the other hand, exhibits extreme behaviour by concentrating the distribution of nearly the entire population within a very narrow range [$0.39$–$0.44$] (c.f., Figure~\ref{fig:KL-divergence}d), pointing an over-generalization of the mean score to all instances. Interestingly, \emph{late-fusion} produces the least biased results for both demographics. However, while aggregating the decisions from different modalities, its average decision gets affected by the extremity of the \emph{visual} modality, leading to over-generalization of the mean score, consequently resulting in higher MAEs (c.f., Figure~\ref{fig:KL-divergence}f). In contrast, \emph{early-fusion} delivers the most accurate predictions with the lowest MAEs (c.f., Figure~\ref{fig:KL-divergence}e) by effectively learning and resolving the unique peculiarities of each modality, such as underestimation, over-generalization, and bimodal distribution, resulting in a shape that resembles the \emph{ground-truth} (c.f., Figure~\ref{fig:KL-divergence}a, ~\ref{fig:KL-divergence}e).

\noindent \textbf{In biased real-world setups (Gender/Ethnicity-Biased):} We observe that the \emph{ground-truth} distributions are not aligned for both demographics (c.f., Figure~\ref{fig:KL-divergence}a). W.r.t. individual modalities (c.f., Figure~\ref{fig:KL-divergence}b -~\ref{fig:KL-divergence}d), we see that the \emph{tabular} modality continues to exhibit underestimation across all demographics, which leads to close alignment of the demographic specific distributions (c.f., Figure~\ref{fig:KL-divergence}b(2) and b(4)). With \emph{textual} modality we notice a misalignment of distribution w.r.t. gender demographics with a favourable skewness for males. However, no such bias is observed w.r.t. ethnicity, indicating a possibility of gender-skewness being much higher than the ethnicity-skewness for the job-related words in the embedded space. Conversely, the \emph{visual} modality demonstrates the most extreme bias for both demographics. Regarding gender, it shows a positive bias towards males, while for ethnicity, it overgeneralizes Asians, discriminates against Blacks, and favours Caucasians. Continuing the trend established in the neutral setup, \emph{Early} fusion consistently mimics the \emph{ground-truth} for both demographics, yielding the lowest MAEs while maintaining fairness. \emph{Late-fusion}, while also following its trend, tends to over-generalize the mean score, resulting in higher MAEs but also higher KL scores. 

In general, leveraging multimodal data can enhance performance and mitigate bias compared to relying on a single modality. However, blindly fusing all modalities may not always yield the best results. For instance, the \emph{tabular} in \emph{gender-biased} setup (c.f., Figure~\ref{fig:KL-divergence}b(2)) and the \emph{textual} in \emph{ethnicity-biased} setup (c.f., Figure~\ref{fig:KL-divergence}c(4)) outperformed both fusion strategies. We hypothesise that \emph{late-fusion} exacerbates biases by independently learning biased models for each modality, cumulatively impacting decision fairness, while \emph{early-fusion} offers greater flexibility and generally yields fairer outcomes with lower prediction error. Dataset diversity and biases may have influenced these findings, highlighting the need to assess robustness across multiple datasets, domains, and fusion strategies. We contemplate that in the future, exploring mid-fusion strategies could enhance fairness and accuracy in decision-making through strategic selection and a combination of modalities.

\section{Conclusions}
    In our study, we used the FairCVdb dataset to investigate the bias implications of \emph{early-} and \emph{late-} fusion strategies in multimodal AI-based recruitment. We assessed biases in gender and ethnicity demographics across both unbiased (neutral) and real-world (gender/ethnicity-biased) setups. Our findings reveal that \emph{early-fusion} closely mimics the ground truth for both demographics, achieving the lowest MAEs by effectively incorporating the unique characteristics of each modality. In contrast, \emph{late-fusion} leads to highly over-generalized mean scores, resulting in higher MAEs. Our evaluation underscores the significant potential of \emph{early-fusion} for applications requiring both accuracy and fairness, providing robust solutions even in the presence of demographic biases. Based on the results, we speculate that mid-fusion strategies may enhance fairness and accuracy by strategically selecting and combining modalities. Exploring these findings across diverse datasets and domains beyond hiring could further broaden the study's impact and relevance. \textbf{Ethics statement:} Understanding the risks of using simulated or synthetic data is crucial for fairness, transparency, and effectiveness in automated hiring processes. 
\begin{acknowledgments}
    This research work is funded by the European Union under the Horizon Europe MAMMOth project, Grant Agreement ID: 101070285. UK participant in Horizon Europe Project MAMMOth is supported by UKRI grant number 10041914 (Trilateral Research LTD). The research is also supported by the EU Horizon Europe project STELAR, Grant Agreement ID: 101070122.
\end{acknowledgments}
\bibliography{main}
\end{document}